\documentclass[12pt]{article}
 
 \newcommand{\n}{\noindent} 
\newcommand{\rf}[1]{(\ref{#1})}
\newcommand{\ba}{\begin{array}} \newcommand{\ea}{\end{array}}
\newcommand{\be}{\begin{equation}} 
\newcommand{\btb}{\begin{tabular}}\newcommand{\etb}{\end{tabular}}
\newcommand{\ee}[1]{\label{#1}\end{equation}}
\newcommand{\bi}{\bibitem} 
 
\newtheorem{thm}{Theorem}[section] 
 
\newcommand{\dss}{\displaystyle}
\newcommand{\bfl}{\begin{flushleft}}\newcommand{\efl}{\end{flushleft}}
\newcommand{\td}{\tilde}

\newcommand{\al}{\alpha} \newcommand{\bt}{\beta}
\newcommand{\g}{\gamma}  \newcommand{\de}{\delta}
\newcommand{\ep}{\epsilon}
\newcommand{\te}{\theta} \newcommand{\De}{\Delta}
  \newcommand{\si}{\sigma}
\newcommand{\La}{\Lambda}
\newcommand{\Si}{\Sigma}
\newcommand{\C}{\mathbb C}  \newcommand{\R}{\mathbb R}

\newcommand{\LCU}{{\cal L}_\uparrow}

\newcommand{\B}{{\bf B}}\newcommand{\E}{{\bf E}} \newcommand{\F}{{\bf F}} 
\newcommand{\0}{{\bf 0}}
\newcommand{\ay}{{\bf a}}

\newcommand{\fy}{{\bf f}} 
 
\newcommand{\jy}{{\mathbf  j}} 
 
\newcommand{\ry}{{\bf r}}
\newcommand{\Ay}{{\bf A}} \newcommand{\By}{{\bf B}}
\newcommand{\Dy}{{\bf D}} 
\newcommand{\Fy}{{\bf F}}\newcommand{\Hy}{{\bf H}}

\newcommand{\Vy}{{\bf V}}

\newcommand{\vy}{{\bf v}}

\newcommand{\hF}{\hat{F}}

\newcommand{\na}{\nabla} \newcommand{\p}{\partial}  
 \newcommand{\we}{\wedge} 
\newcommand{\ra}{\rightarrow} \newcommand{\ol}{\overline}

\newcommand{\vo}{\mbox{vol}} 
\newcommand{\di}{\mbox{div}}\newcommand{\Di}{\mbox{Div}}
\newcommand{\lo}{\mbox{lor}} 
\newcommand{\diag}{\mbox{diag}} 
 
\newcommand{\cu}{\mbox{curl}}

\usepackage{amssymb}
\usepackage{graphicx, epstopdf,}
\usepackage{hyperref}
\usepackage{breakurl}

\begin{document} 

\title{On the relativistic unification \\ of electricity and magnetism}
\normalsize 
\author{Marco Mamone Capria\footnote{Corresponding author: University of Perugia, Dipartimento di Matematica -- via Vanvitelli 1, 06123 Perugia, Italy; \texttt{mamone@dmi.unipg.it}}\, and Maria-Grazia Manini} 
\maketitle 

\begin{quote}\small 
{\bf  Abstract} 
The unification of electricity and magnetism achieved by special relativity has remained for decades a model of unification in theoretical physics. 
We discuss the relationship between electric and magnetic fields from a classical point of view, and then examine how the four main relevant authors (Lorentz, Poincar\'e, Einstein, Minkowski) dealt with the problem of establishing the transformation laws of the fields in different inertial systems. We argue that Poincar\'e's derivation of the transformation laws for the potentials and the fields was definitely less arbitrary than those of the other cited authors, in contrast with the fact that here, as in other instances, Poincar\'e's contribution to relativity was belittled by authoritative German physicists in the first two decades. In the course of the historical analysis a number of questions which are of contemporary foundational interest concerning relativistic electromagnetism are examined, with special emphasis on the role of potentials in presentations of electromagnetism, and a number of errors in the historical and foundational literature are corrected.

{\bf Keywords} classical electromagnetism, special relativity, potentials, history of relativity 

\end{quote}\normalsize

\tableofcontents

\begin{quote}\small 
{\sc Lorentz} a cherch\'e alors \`a la completer et \`a la modifier de fa\c{c}on \`a la mettre en concordance parfaite avec ce postulat. C'est ce qu'il a r\'eussi 
\`a faire dans son article intitul\'e ``Electromagnetic phenomena in a system moving with any velocity smaller than that of light'' [...] [{\sl H. Poincar\'e}, 1906, \cite{poi06}, p. 129]\footnote{``Then {\sc Lorentz}  tried to complete [his contraction hypothesis]  and to modify it in order to make it agree perfectly with this postulate [the postulate of relativity]. It is what he succeeded in doing in his article `Electromagnetic phenomena in a system moving with any velocity smaller than that of light' [...].''}

F\"ur jene urspr\"unglichen Gleichungen ist die Kovarianz bei den Lorentz-Transforma\-tionen eine rein mathematische Tatsache, die ich das Theorem der relativit\"at nennen will. [...] {\sc H. A. Lorentz} hat das Relativit\"atstheorem gefunden und das Relativit\"atspostulat geschaffen [...] [{\sl H. Minkowski}, 1908, \cite{min08a}, pp. 54, 55]\footnote{``The covariance of these fundamental equations under the Lorentz transformations is a purely mathematical fact [...] {\sc H. A. Lorentz} has found out the `Relativity theorem' and has created the `Relativity-postulate' ''.}

En effet, pour certaines des grandeurs physiques qui entrent dans les formules, je n'ai indiqu\'e la transformation qui convient le mieux. Cela a \'et\'e fait par {\sc Poincar\'e} et ensuite par {\sc M. Einstein} et {\sc Minkowski}. [{\sl H. A. Lorentz}, 1915, \cite{lor15}, p. 295]\footnote{``In fact, for some of the physical quantities entering into the formulas, I have not indicated the best suitable transformation. This has been done by {\sc Poincar\'e}, and later by {\sc M. Einstein} and {\sc Minkowski}.''}
\end{quote} \normalsize

\section{Introduction}

The unification of electricity and magnetism achieved by special relativity has remained for decades a model of unification in theoretical physics. The melting together of  two 3-dimensional entities, the electric and magnetic fields, into a single 4-dimensional entity has given an acceptable mathematical meaning to the very idea of `unifying' different interactions. On the other hand it is clear that the problem of `unification' of interactions admits of more than one interpretation and solution, a fact which has been brought home by the whole development of theoretical physics in the 20th century. 

After laying down the basic definitions and formulas (\S 2), we discuss the relationship between classical physics and electromagnetism, with emphasis on unification of the electric and the magnetic fields {\sl before} special relativity. We focus on Hertz's theory and Poincar\'e's reaction to it, but also deal with some modern neo-classical contributions (\S 3).  

Section 4 is the core of the paper, and contains an analysis of the quite different steps taken by the founders of special relativity (Lorentz, Poincar\'e, Einstein, and Minkowski) in order to reach a central result of the theory: the Lorentz covariance of the Maxwell equations -- which may be called `the fundamental theorem of relativistic electromagnetism' .\footnote{The Lorentz -- or rather Poincar\'e -- {\sl covariance} of the Maxwell equations is often called in the literature `the Lorentz {\sl invariance}' of those equations, which is certainly correct if one thinks in intrinsic terms rather than in components (cf. \S 4.4, \S 5).}

We examine the logical structure of their arguments, and the differences between them. In particular, we evaluate the way each author derived the {\sl transformation laws for the electric and magnetic fields}, since such laws define uniquely the nature of the new 4-dimensional algebraic entity in which the 3-dimensional fields are `assembled' -- and it is the transparency and unambiguity of the procedure introducing  this entity (explicitly or implicitly) which is the test of the unification. Our four authors all got the same transformation laws for the fields, but -- contrary to what seems to have been the attitude of  most historians of special relativity (see e.g. \cite{mi81}, \cite{tor83}, \cite{za89}),\footnote{An extreme case is provided by the well-known book by Stanley Goldberg (\cite{gol84}), which has a very detailed analytical index (pp. 485-94), where, however, neither `Maxwell's equations' nor `Maxwell, J. C.' are listed (in fairness, Maxwell's equations are mentioned at least once in this book, at p. 105: ``the basic laws of electroamgnetic theory, commonly referred to as Maxwell's equations'', and Maxwell himself at p. 433).} -- we think it important to analyze just {\sl how} those laws are obtained by each author, and how satisfactory is each derivation. 

In the following section (\S 5) we provide an outline of the modern presentation of relativistic electromagnetism, emphasizing the influence of the authors we have discussed and commenting on some very recent criticism of the relativistic unification of electricity and magnetism. The historical and the foundational threads in our treatment find their common ground in the long-standing conflict between two different approaches to classical electromagnetism, the one based on fields and the other based on potentials (\S 4.2.2, \S 5.1).      

In the course of our inquiry several foundational issues are briefly discussed, and some errors in the foundational and historical literature are corrected. It may be helpful to some readers to be warned that the historical scope of our article is limited, since we focus on the unification issue only, and leave aside several topics which are treated in the papers we analyze (in particular \cite{lor04}, \cite{poi06}, \cite{ei05}, \cite{min08a}), although they may be important (or even {\sl more}  important) in assessing the respective contributions of those authors to the relativity revolution. Nonetheless, from an historical point of view, the way our authors succeeded or failed in giving an unambiguous translation of classical electromagnetism in the new framework is clearly a relevant piece of evidence for a correct attribution, {\sl and as such it was perceived by some of the main actors}. So our paper is meant to counterbalance the tendency among historians of the priority controversy concerning special relativity to completely disregard this aspect of the question.\footnote{See, for instance, the defense of Einstein's priority contained in \cite{cerf06}. The recognition due to Poincar\'e has been too long obfuscated by a widespread overreaction to Whittaker's chapter on special relativity in his famous history of electromagnetism (`The relativity theory of Poincar\'e and Lorentz') (\cite{whi53}, vol. II, pp. 27-77). While the reconstruction presented in that chapter is in several ways unsatisfactory, the stress on the importance and primacy of Poincar\'e's contribution (to be more carefully separated from Lorentz's than Whittaker did) is on the whole warranted, and our analysis strengthens it in one point that, curiously, was ignored by both Whittaker and his critics.}

In the final section (\S 6) we shall show that an inflation of the role of Lorentz and Einstein to the detriment of Poincar\'e's was a main feature of the historical outlines contained in some important articles and treatises on relativity as regards the specific contributions of these scientists to the unification issue. We think this issue to be of interest to both physicists and historians of contemporary physics, insofar as it emphasizes the need of double-checking before taking attributions of results and citations of authors at face value.     

Our treatment adopts throughout a unified modern notation for the sake of clarity and because the authors we are studying used different conventions (for instance while Poincar\'e and Minkowski set $c =1$, Lorentz and Einstein did not; Lorentz and Minkowski used modern vector notation, Poincar\'e and Einstein did not; etc.). We have also endeavoured to provide enough details to make it possible to use fragments of this story in a modern introduction to relativistic electromagnetism, and to encourage both scientists and historians of ideas to base their own image of the past on the primary sources.

\section{Preliminaries}

In this section we shall list some of the basic notions that will occur in the following, and fix our conventions. 

The form of the {\sl Maxwell equations} (which in fact were established as a system by O. Heaviside -- \cite{hunt}, pp. 245-7 -- and, more similar to the modern presentation, by H. Hertz -- \cite{dar}, p. 237)  in the aether (in MKSQ units) we shall adopt is: 

\be \left\{\ba{rcl} \di \E &=& \dss\frac{\rho}{\ep_0} \\ [8pt] \di \B &=& 0 \\ [8pt]  \cu \E &=& -\dss\frac{\p\B}{\p t} 
\\ [8pt]  \cu\B &=& \mu_0 \jy + \dss\frac{1}{c^2}\frac{\p\E}{\p t} \ea\right. ,\ee{max}

\n
where $\E$ is the electric field, $\B$ is the magnetic induction, and the charge and current density functions $\rho$ and $\jy$ are related by the continuity equation (which can also be viewed as a consequence of \rf{max}$_1$ and \rf{max}$_4$):

\be \di \jy + \frac{\p\rho}{\p t} =0. \ee{ce} 

\n 
The constants $\ep_0$ and $\mu_0$ are related to the speed of light in the aether\footnote{All authors we shall analyse, except of course for Einstein, freely referred quantities and equations to the `aether'; for instance, the first part of Minkowski's article \cite{min08a} bears the title ``Betrachtungen des Grenzfalles \"Ather'', and section 2 is entitled: ``Die Grundgleichungen f\"ur den \"Ather''.} $c$ by the equation \(c = (\ep_0 \mu_0)^{-1/2}\). An alternative formulation, which we shall employ only in the case of Minkowski (though it is also the one used by Lorentz \cite{lor04}), introduces the electric displacement  $\Dy$ and the magnetic field $\Hy$, defined in an isotropic medium with permittivity $\ep$ and permeability $\mu$ as

\[ \Dy = \ep \E, \; \Hy = \frac{1}{\mu} \By, \]

\n
to give the mathematically equivalent system: 

\be \left\{\ba{rcl} \di \Dy &=& \rho \\ [8pt] \di \Hy &=& 0 \\ [8pt]  \cu \Dy &=& -\dss\frac{1}{c^2} \frac{\p\Hy}{\p t} 
\\ [8pt]  \cu\Hy &=& \jy + \dss\frac{\p\Dy}{\p t} \ea\right. .\ee{max_twin}

The Maxwell equations can be expressed by means of two auxiliary functions, the magnetic (vector) and electric (scalar) potentials $\Ay$ and $\psi$, if we put

\be \left\{\ba{rcl} \E &=& -\dss\frac{\p \Ay}{\p t} - \na \psi \\ [8pt] \By &=& \cu\Ay .\ea \right. \ee{fields}

\n
By introducing one of the main characters in our story, the d'Alembert operator:

\[ \Box \equiv \sum_{\al=1}^3 \frac{\p^2 }{\p (x^\al)^2} -\frac{1}{c^2}\frac{\p^2 }{\p t^2}, \] 

\n
and by using standard formulas of vector calculus, we obtain:

\[ \Box \Ay = -\cu\B +\frac{1}{c^2}\frac{\p \E}{\p t} + \na (\di\Ay + \frac{1}{c^2} \frac{\p \psi}{\p t}) \]

\[ \Box \psi = -\di\E - \frac{\p}{\p t} (\di\Ay + \frac{1}{c^2} \frac{\p \psi}{\p t}). \]

\n
Therefore the Maxwell equations are equivalent to

\be \Box \Ay = - \mu_0 {\mathbf j}, \; \Box \psi = - \frac{\rho}{\ep_0} \ee{maxp} 

\n
subject to the Lorenz\footnote{This condition, which is often attributed to Lorentz (e.g. \cite{mi73}, p. 250; \cite{za89}, p. 177) -- a bias which incidentally provides a good example of the Matthew effect -- was introduced by the Danish physicist Ludwig V. Lorenz in 1867 (\cite{lor67a}, \cite{lor67b}; \cite{whi53}, I, p. 269; cf. \cite{ja08}, pp. 704-7). We shall come back to this issue in \S 4.1.1.} (gauge) condition:

\be \di\Ay + \frac{1}{c^2}\frac{\p \psi}{\p t} = 0, \ee{lgc}

\n
which was shown by Lorenz (\cite{lor67b}, p. 294) to be satisfied by the retarded potential solutions of  \rf{maxp}:

\be \left\{\ba{rcl}  \Ay (\ry_0, t_0)  &=& \dss\frac{\mu_0}{4\pi}\int_{\R^3} \frac{[\jy]}{s} \vo_3, \\ [8pt]   
\psi (\ry_0, t_0) &=& \dss \frac{1}{4\pi\ep_0}\int_{\R^3}\frac{[\rho]}{s} \vo_3,\ea\right. 
\ee{retp}

\n
where the square brackets stand for the `retarded' functions relative to the base point $(\ry_0, t_0)$:

\[ [f] (\ry, t) : = f(\ry, t_0- s/c), \; \mbox{with}\; s: = | \ry -\ry_0|. \]

\n
We shall see that potentials play a crucial role in our reconstruction of the differences between Lorentz, Poincar\'e and Einstein.

The special conformal Lorentz transformation (notice that the name Poincar\'e gave it was simply `Lorentz transformation' -- \cite{poi05b}, p. 1505) is 

\be \left\{\ba{rcl} x'^1 &=& l\al (x^1 - Vt), \\ x'^2 &=& l x^2 \\ x'^3 &=& l x^3 \\ t'&=& l\al (t - Vx^1 /c^2) \ea\right. \ee{cslt}

\n
where

\[ \al = \frac{1}{\sqrt{1-\bt^2}}, \]

\n
and $l$ is  a constant positive factor. The inverse transformation of \rf{cslt} is

\be \left\{\ba{rcl}  x^1 &=& \dss \al l^{-1} (x'^1 + Vt'), \\ [6pt] x^2 &=& l^{-1} x'^2 \\ [6pt] x^3 &=& l^{-1} x'^3 \\ [6pt] t &=& \al l^{-1} (t' + Vx'^1 /c^2) \ea\right. \ee{cslt_inv}

The transformation \rf{cslt} first appeared in print in Woldemar Voigt's article of 1887 on the Doppler effect, with the special choice $l = \al^{-1}$ (equation (10) of \cite{voi}, p. 45)\footnote{An English translation and commentary is provided in \cite{eh01}.}, that is as:

\be \left\{\ba{rcl} x'^1 &=& x^1 - Vt, \\ x'^2 &=& \sqrt{1-\bt^2} x^2 \\ x'^3 &=& \sqrt{1-\bt^2} x^3 \\ t'&=& t - Vx^1 /c^2 \ea\right. .\ee{voi}

\n
Voigt presented the transformation, correctly,  as such as to leave invariant the wave equation in free space. 

We shall denote by $\phi$ and $\phi'$ the `unprimed' and the `primed' coordinate systems (formally they can be seen as functions from the space of events to $\R^4$, and the systems \rf{cslt}, \rf{voi} etc. as their transition function $\phi'\circ\phi^{-1}$).

The following formulas are a direct consequence of \rf{cslt}:

\be \left\{\ba{rcl} \dss\frac{\p}{\p x'^1} &=& \dss\frac{\al}{l} (\frac{\p}{\p x^1} + \frac{V}{c^2} \frac{\p}{\p t})
\\ [8pt] \dss\frac{\p}{\p x'^2} &=& \dss\frac{1}{l} \frac{\p}{\p x^2}
\\ [8pt] \dss\frac{\p}{\p x'^3} &=& \dss\frac{1}{l} \frac{\p}{\p x^3}
\\ [8pt] \dss\frac{\p}{\p t'} &=& \dss\frac{\al}{l} (\frac{\p}{\p t} + V \frac{\p}{\p x^1}) \ea\right. ,\ee{par} 

\n
and will be frequently used in the following.

\section{Electromagnetism and Galilean invariance}

Before considering the `relativistic' developments, let us examine how Maxwellian electromagnetism fits (or rather fails to fit) into the classical framework. In classical physics the basic condition for the fields is that they submit to whatever transformations are required to ensure the Galilean covariance of the {\sl force} expression. Let $\phi$ be a Galilean coordinate system. Since the electric force $\F_e$ on a charge $q$ under an electric field $\E$ is

\be  \F_e = q \E \ee{efor}

\n
while the magnetic force $\F_m$, if $q$ is moving with a velocity $\vy$ under a magnetic field $\B$, is:

\be \F_m = q\vy\we\B, \ee{mfor}  

\n
it follows, by the superposition principle, that the total force on the charge (if we assume that no other forces are acting on it) is the {\sl Lorentz force law}:

\be \F_L = \F_e + \F_m = q (\E + \vy\we \B). \ee{lfor}

Now let $\phi'$ be another Galilean coordinate system, related to $\phi$ by any transformation of the form

\[ \left\{\ba{rcl} \ry' &=& S(\ry - t\Vy) \\ [4pt] t' &=& t+a \ea\right. \]

\n
where $S\in SO(3)$, $a\in \R$, and $\Vy\in \R^3$ is the velocity of $\phi'$ with respect to $\phi$. We shall denote by $\Vy'$ the reciprocal velocity, that is \(\Vy'= - S\Vy\).

By requiring that $\F'_L = S\F_L$ we obtain 

\be \left\{\ba{rcl} \E' &=& S(\E + \Vy \we\B) \\ [4pt] \B' &=& S\B, \ea\right. \ee{gft}

\n
which shows that classically there must be 1) a link between the electric and the magnetic fields; 2) a strong asymmetry between them, as the former can be destroyed by a coordinate change, while the second cannot. 

At this point the problem of whether the Maxwell equations are covariant or not under the Galileo group has acquired a perfectly definite meaning. Suppose $\phi$ is aether-fixed, so that we can apply in it the Maxwell equations. Consider the homogeneous pair first, that is the pair of equations \rf{max}$_2$ and \rf{max}$_3$ which are the same whether $\rho$ and $\jy$ are zero or not. The Gauss law for the magnetic field is clearly Galileo covariant, since

\[ \di'\B' = \na'\cdot\B' = S\na\cdot S\B = \na\cdot \B = \di\B =0.\]

Instead for the Faraday-Henry law one obtains in $\phi'$ a quite different law:

\be \cu'\E' = -\frac{\p\B'}{\p t} - (\Vy\cdot\na)\B' . \ee{fh0}

\n
Let us turn now to the inhomogeneous pair of \rf{max}. First of all we have to find out how the density functions transform, and this is rather easy to do (cf. \cite{bm}, p. 794):

\be \rho' (\ry', t') = \rho (S^T \ry'+ t'\Vy, t'), \; \jy' (\ry', t') = \rho'\vy'= \rho' S (\vy - \Vy).
\ee{den}

\n
A simple computation shows that these functions satisfy the continuity equation \rf{ce}. 

On the other hand, the following computation:

\[ \di'\E'= \frac{\rho'}{\ep_0} -\cu\B\cdot\Vy  = \frac{\rho'}{\ep_0} - (\mu_0\jy + \frac{1}{c^2}\frac{\p \E}{\p t})\cdot\Vy, \]

\n
shows that the Gauss law for the electric field is not valid in $\phi'$. Finally consider the Maxwell-Amp\`ere equation:

\[  \cu'\B' = S\na\we S\B = S(\na\we\B) = \dss S(\mu_0\jy + \frac{1}{c^2}\frac{\p \E}{\p t}). \]

\n
This is equivalent to:

\be  \cu'\B'  = \dss\mu_0\jy '+ \frac{1}{c^2}\frac{\p \E'}{\p t} +\frac{1}{c^2}(\Vy'\we\frac{\p\B'}{\p t} -\frac{\rho'}{\ep_0}\Vy'),
\ee{am0}

\n
so even the fourth Maxwell equation is not Galileo covariant.  It follows that if we wish to have a set of equations which are Galileo covariant we must modify the Maxwell equations. This is indeed what has been attempted at the end of the 19th century.

\subsection{Hertz's compromise and Poincar\'e's criticism}

In a paper of 1890, bearing a title strikingly similar to that of Einstein's 1905 relativity article, that is ``On the fundamental equations of the electrodynamics of moving bodies'' (\cite{hertz90}), Hertz expressed the Gauss laws in the more symmetrical form

\be \di \E = \frac{\rho}{\ep_0}, \; \; \di \B = \mu_0 \rho_m, \ee{gauss_her} 

\n
thus including the possibility that magnetic monopoles might exist. He then modified the other two equations by substituting the partial time derivatives by convective derivatives:

\be \frac{d}{dt} := \frac{\p}{\p t} + \Vy\cdot\na . \ee{der_h} 

It is clear that by introducing this derivative, \rf{fh0}  can be re-written as:

\be \cu'\E' = -\frac{d\B'}{d t} . \ee{fhh}

\n
Since obviously for an aether-fixed coordinate system $\dss\frac{d}{d t} = \frac{\p}{\p t}$, this version of the Faraday-Henry law generalizes the aether-fixed law. On the other hand, the Amp\`ere-Maxwell equation must be {\sl postulated}  to be: 
 
\be \cu'\B' = \mu_0 \jy' + \dss\frac{1}{c^2}\frac{d\E'}{dt}, \ee{mah}

\n
which is different from \rf{am0}. According to Hertz, ponderable bodies drag the aether within them totally, though he concedes that this is more a working hypothesis than an established empirical truth. It is not clear which assumptions concerning the transformation of the fields Hertz had in mind, but \rf{gauss_her} suggests that he thought that $\E$ and $\B$ were invariant (up to a rotation), which is what T. E. Phipps will formally advance a few years ago, as we shall see.

An analysis of Hertz's theory was provided by Poincar\'e in a detailed survey of contemporary theories of electromagnetism published five years later (\cite{poi95}). Under the assumptions

\[ \jy = \0, \; \rho = 0, \; \rho_m =0 \]

\n
in a medium with refraction index $n$ and light velocity $c_n : = c/n$ one has:

\[\left\{\ba{rcl} \cu\E &=&  -\dss\frac{d\B}{dt} \\ [6pt] \cu\B &=& \dss\frac{1}{c_n^2} \frac{d\E}{dt}.\ea\right. \]

\n
Now if $\Vy = const.$, a standard computation gives:

\[ \De\B = \frac{1}{c^2_n} (\frac{\p^2\B}{\p t^2} + 2(\Vy\cdot\na)\frac{\p\B}{\p t} + (\Vy\cdot\na)(\Vy\cdot\na)\B), \]

\n
and if, with Poincar\'e, we consider \(\Vy = (0,0,V)\), we get:  

\[ \De\B = \frac{1}{c_n^2} (\frac{\p^2\B}{\p t^2} + 2V\frac{\p^2\B}{\p z \p t} + V^2 \frac{\p^2\B}{\p z^2}), \]

\n
which corresponds to a wave velocity

\[ w = \pm c_n - V .\]

\n
Such a formula, however, implies a complete dragging of the aether by the medium, in contradiction with Fizeau's experiment, according to which:

\be w = \pm c_n - V (1-\frac{1}{n^2}).\ee{fiz}

\n
This is the main reason Poincar\'e found Hertz's theory unsatisfactory, although he rated it high, since it was ``the only theory compatible with the principle of conservation of electricity and magnetism and the principle that action and reaction are equal'' (\cite{poi95}, p. 409).

\subsection{Modern proposals and arguments}

Notice that to start with the right foot (that is, the force law) is essential. For instance Jammer and Stachel (\cite{js80}) introduce an {\sl ad hoc} definition of the field transformation, which in our units is (by taking, as they do, $S=I_3$):

\be \left\{\ba{rcl} \E' &=& \E \\ [4pt] \B' &=& \B +\frac{1}{c^2}\Vy\we\E , \ea\right. 
\ee{gft_js}

\n
Under \rf{gft_js} the Gauss law is clearly preserved, and the same is true for the Amp\`ere-Maxwell equation (using the correct transformation laws for the density functions, that is \rf{den}); however in the case of the Gauss law for the magnetic field one obtains 

\[ \di'\B'= -\frac{1}{c^2}\Vy\cdot \cu\E \]

\n
so the authors' proposal (which is not advanced seriously, though, but only as ``an historical fable with a pedagogical moral'') is to modify Faraday-Henry law into

\be \cu\E = \0, \ee{fhjs} 

\n
which with one blow would give the covariance of \rf{max}$_2$ and, of course, \rf{max}$_3$. However with the transformation \rf{gft_js} the correct force law would have to be not the standard Lorentz force $\F_L$, but  

\be \F_{JS} = q (c\B +\frac{1}{c}\vy\we \E), \ee{jsfor}

\n
which does not make physical sense, for the straightforward reason that a charge is not acted upon by a magnet if they are relatively at rest (in the aether). However, this was first proven experimentally by Faraday in 1831, and these authors are making a historical thought experiment on what would have happened ``if Maxwell had worked between Amp\`ere and Faraday''. 

A different approach, which follows Hertz in substituting the partial with a special convective time derivative in Maxwell equations, is that of T. E. Phipps; his equations are therefore:

\be \left\{\ba{rcl} \di \E &=& \dss\frac{\rho}{\ep_0} \\ [8pt] \di \B &=& 0 \\ [8pt]  \cu \E &=& -\dss\frac{d\B}{d_P t}, 
\\ [8pt]  \cu\B &=& \mu_0 \jy + \dss\frac{1}{c^2}\frac{d\E}{d_P t} \ea\right. ,\ee{max_p}

\n
\n
where the convective derivative introduced by Phipps is

\be \frac{d}{d_P t} := \frac{\p}{\p t} + \Vy_d\cdot\na, \ee{der_p} 

\n
and $\Vy_d$ is the velocity of the detector, that is the measuring device.  However, for the formal machinery to work, Phipps must require for the fields the following transformation laws:

\be \E' = S\E, \B' = S\B, \ee{gft_p}

\n
in other words they must be invariant (up to rotation). Needless to say, also this proposal contradicts the Lorentz force law; the author admits the consequence, but he is willing to accept it.\footnote{``To be sure, there are many twists of the actual historical process we have not attempted to bring into our discussion. For instance, the Lorentz force law, which emerged contemporaneously with Hertz's theory, fits with covariance rather than invariance. But present-day empirical evidence is mounting heavily against the Lorentz force law and in favor of Amp\`ere original law of forces between current elements [$\ldots$], which honored Newton's third law but not covariance or space-time symmetry'' (\cite{phipps93}).} 

Another recent revival of Hertz theory is presented by R. T. Cahill in \cite{cah08}, which argues in favour of the anisotropy of light that this theory predicts.

\subsection{Electromagnetism and Galilean space-time}

In this section we have seen some of the theoretical difficulties that prevented a natural inclusion of electromagnetism into Newtonian physics, and some of the proposed solutions. We have emphasized that any transformation law of the electric and magnetic fields from an aether coordinate sytem to an inertial system  necessarily involves, to be compatible with the Lorentz force law,  a weakening of the mutual independence of the fields -- therefore any such law `unifies' them. At a deeper level this can be explained by reference to the underlying space-time structure of the Newtonian force principle, the Galilean structure defining what may be called the Galilean space-time. It is important to emphasize that space-time, as something distinct from absolute space and absolute time taken together, is already present in classical physics.

\section{Electromagnetism and the rise of special relativity}

We begin with the basic chronology of the rise of relativistic electromagnetism (\cite{whi53}, \cite{mi81}, \cite{dar}, \cite{ora}). The main historical articles we shall examine from the viewpoint of the unification of the electric and the magnetic fields are: 

\begin{itemize}
\item 
Hendrik A. Lorentz's ``Electromagnetic Phenomena in a System Moving with any Velocity less than that of Light'' (1904: \cite{lor04}), 

\item
Henri Poincar\'e's two articles with the same title,``On the dynamics of the electron'' (1905: \cite{poi05b}, 1906: \cite{poi06}), 

\item 
Albert Einstein's ``On the Electrodynamics of Moving Bodies'' (1905: \cite{ei05}),  

\item
Hermann Minkowski's ``The principle of relativity'' (1907: \cite{min15})  ``The fundamental equations for the electromagnetic processes in moving bodies'' (1908: \cite{min08a}). 
\end{itemize}

Note that  \cite{poi05b} is a short note which was read at the session of 5 June 1905 of the French Academy of Sciences, while the long article \cite{poi06} (of which \cite{poi05b} is an outline) is dated by its author as ``July 1905'' and was read at the session of 23 July of Palermo's Mathematical Circle.  About half-way between these two dates, and precisely on 30 June, the German journal {\sl Annalen der Physik} received Einstein's manuscript.\footnote{In analysizing the articles listed above we normally use the present tense (`Lorentz writes' etc.) in order to stress that we are focusing on the text under examination rather than on the precise historical circumstances of when, where and even who materially wrote the texts we are analysing and/or quoting from.}    

\subsection{Lorentz, 1904}

In his last statement before 1905, Lorentz writes the full system of Maxwell equations (in the form \rf{max_twin}), together with the expression for the force. He then assumes that the same equations are true in a coordinate system $\ol{\phi}$ which moves with respect to (aether-based) $\phi$ with constant velocity $\Vy$ directed along the $x ^1$-axis, which means, as he put it, that the velocity of ``a point of an electron'' is 

\be\ol{\vy} = \vy + \Vy. \ee{vellor}

\n
Although Lorentz does not write it down, the coordinate change involved is: 

\be \left\{\ba{rcl} \ol{x}^1 &=& x^1 - Vt \\ \ol{x}^2 &=&  x^2 \\ \ol{x}^3 &=& x^3 \\ \ol{t} &=& t \ea\right., \ee{gal}

\n
which implies that, in $\ol{\phi}$, \rf{max} can be re-written as

\be \left\{\ba{rcl}\ol{\di} \E &=& \dss\frac{\rho}{\ep_0} \\ [8pt] \ol{\di} \B &=& 0 \\ [8pt]  \ol{\cu} \E &=& -\dss (\frac{\p}{\p t} - V \frac{\p}{\p x^1})\B  
\\ [8pt]  \ol{\cu}\B &=& \mu_0 \jy + \dss\frac{1}{c^2}(\frac{\p}{\p t} - V \frac{\p}{\p x^1})\E \ea\right. ,\ee{max_1}

\n
where the overlined operators must be understood with respect to $\ol{\phi}$. Of course \rf{vellor} is the Galilean addition law for velocity, so the outcome of Lorentz's argument cannot be but an {\sl approximation} to what will come to be known as `Lorentz invariance'. Lorentz then introduces a new coordinate system $\phi'$, related to $\ol{\phi}$ by the equations:

\be \left\{\ba{rcl} x'^1 &=& \al l \ol{x}^1  \\ x'^2 &=&  l\ol{x}^2 \\ x'^3 &=& l\ol{x}^3 \\ t' &=& \dss\frac{l}{\al} \ol{t} - \al l\frac{V}{c^2}\ol{x}^1 \ea\right. . \ee{lor}

It is interesting to remark that the Lorentz transformations do not appear {\sl explicitly} in this article, although by inserting \rf{gal} into \rf{lor} and taking account of the simple identity

\[ \frac{\ol{t}}{\al} - \al \frac{V}{c^2} \ol{x}^1 = \al (t - \frac{V}{c^2} x^1) \]

\n
one obtains \rf{cslt}. Then Lorentz introduces {\sl by definition} the transformation laws from $\phi$ to $\phi'$ for the fields:  

\be \left\{\ba{rcl} E'^1 &=& \dss\frac{1}{l^2} E^1  \\ [4pt]              
E'^2 &=&  \dss\frac{\al}{l^2} (E^2 -V B^3)  \\ [4pt]
E'^3 &=& \dss\frac{\al}{l^2}(E^3+V B^2)  \ea \right. , 
\left\{\ba{rcl}   B'^1 &=& \dss\frac{1}{l^2} B^1 \\ [4pt] B'^2 &=&  \dss\frac{\al}{l^2} (B^2 + \frac{V}{c^2} E^3)  
\\ [4pt]  B'^3 &=& \dss\frac{\al}{l^2} (B^3 -\frac{V}{c^2} E^2), \ea\right.  
\ee{ftlor}  

\n
and he assumes that $l = l(V) = 1 + O(\bt^2)$. At this point Lorentz writes (\cite{lor04}, p. 813):

\begin{quote}
The variable $t'$ may be called the ``local time''; indeed, for $\al = 1$, $l =1$, it becomes identical with what I have formerly understood by this name.
\end{quote}

\n
This passage is crucial to understand how far Lorentz is from a relativity theory, even though his formalism is largely the same as the one that will be found in Poincar\'e's and Einstein's 1905-06 articles. In fact he introduces the transformation laws of the charge density and of the current velocity by the following 
position:

\be \rho'_L= \frac{1}{\al l^3}\rho, \;  \vy'_L= (\al^2 \ol{v}^1, \al \ol{v}^2, \al \ol{v}^3). \ee{curlor}  

\n
From \rf{curlor} it follows, by using \rf{par}, that the Maxwell equations in the primed coordinate system are:

\be \left\{\ba{rcl} \di' \E' &=& \dss(1-\frac{Vv'^1_L}{c^2})\frac{\rho'_L}{\ep_0} \\ [8pt] \di' \B' &=& 0 \\ [8pt]  \cu' \E' &=& -\dss\frac{\p\B'}{\p t'} 
\\ [8pt]  \cu'\B' &=& \mu_0 \rho'_L \vy'_L  + \dss\frac{1}{c^2}\frac{\p\E'}{\p t'} \ea\right. ,\ee{max_lor}

\n
and this is almost, but not quite, the same as \rf{max} with `primed' quantities. Clearly the origin of this nonrelativistic result is in the second of \rf{curlor}, which is a compromise aiming at retaining as much as possible of the Galilean velocity addition law \rf{vellor}. Apart from this `classical' constraint, the natural definition for the charge density would have been that making the right-hand side of the Gauss law for the electric field {\sl exactly equal}  to $\rho'/\ep_0$. And of course this would have been the obvious decision in case Lorentz had recognized the correct velocity addition law coming from `his' coordinate transformations -- that is, if he had taken seriously \rf{cslt}. In fact from the right-hand side of Gauss law for the electric field one would obtain covariance by simply positing: 

\[ \rho' = \dss(1-\frac{Vv'^1_L}{c^2})\frac{\rho'_L}{\ep_0} = \frac{\al}{l^3} (1- \frac{Vv^1}{c^2})\rho  \]

\n
and from the Amp\`ere-Maxwell equations one would get for the current density

\[ j'^1 = \rho'(\frac{v'^1 - V}{1-Vv'^1/c^2}), \; j'^2 = \rho'(\frac{v'^2}{\al (1-Vv'^1/c^2)}),\; j'^3 = \rho'(\frac{v'^3}{\al (1-Vv'^1/c^2)}).  \]

\n
These are the correct relativistic formulas found by both Poincar\'e and Einstein -- each one following a different route, as we shall see. 

\subsubsection{Potentials and the Lorenz condition}

Lorentz's lack of care for formal invariance either as a heuristic device or as a theoretical constraint (principle of relativity) is also clear from his handling of the potentials. He writes down the equations

\[ \Box' \Ay' = -\mu_0 \jy'_L, \Box'\psi'= -\frac{\rho'_L}{\ep_0}, \; \mbox{where}\; \jy'_L = \rho'_L \vy'_L, \]

\n
and then, without any justification, the expression of the fields appear, in terms of $\Ay'$ and $\psi'$ (formulas (13) and (14)  of \cite{lor04}):

\be \left\{\ba{rcl} \E' &=& -\dss\frac{\p \Ay'}{\p t'} - \na' \psi'  +V\na' A'^1 \\ [6pt]  \By' &=& \cu'\Ay' , \ea\right.  \ee{fields_lor}

\n
Clearly \rf{fields_lor} is not the primed version of \rf{fields}. In his paper \cite{poi06} (p. 134) Poincar\'e will say that his transformation laws for the potentials ``are remarkably different from those of Lorentz'', but in fact neither \cite{lor04} nor the contemporary Encyclopedia article \cite{lor04-} (to whose sections 4,5,10 Lorentz refers) provide explicit expressions for $A'^1, A'^3, A'^3, \psi'$, so the readers are left with an exercise. The likely procedure omitted by Lorentz is the following. Let us consider $A'^1$:

\[ \Box'A'^1 = -\mu_0 j^1_L = -\frac{\mu_0\al}{l^3}\rho (v^1 -V). \]

\n
Since $\Box'= l^{-2} \Box$ (a formula which is not stated in \cite{lor04}), we have

\be \Box A'^1 = -\frac{\mu_0\al}{l}\rho v^1 + \frac{\mu_0\al}{l}\rho V. \ee{a-uno}

\n
On the other hand from

\[ \Box A^1 = -\mu_0 \rho v^1, \; \Box \psi = -\frac{\rho}{\ep_0} \]

\n
one obtains that \rf{a-uno} is satisfied by 

\[ A'^1 = \dss\frac{\al}{l} (A^1 - \frac{V}{c^2} \psi) \]

\n
and the same argument leads to the other components of $\Ay'$ and to $\psi'$:

\be \left\{ \ba{rcl} A'^1 &=& \dss\frac{\al}{l} (A^1 - \frac{V}{c^2} \psi)
\\  A'^2 &=& \dss\frac{A^2}{l}
\\  A'^3 &=& \dss\frac{A^3}{l} 
\\ \psi' &=& \dss\frac{1}{l\al} \psi
\ea\right. \ee{pot_l}
 
Now if in $\phi$ the potentials are related to the fields by \rf{fields}, {\sl the same formulas cannot apply in $\phi'$}. In fact:

\[ \ba{rcl} E'^1 &=& \frac{1}{l^2} E^1 = -\dss \frac{1}{l^2} (\frac{\p\psi}{\p x^1} + \frac{\p A^1}{\p t}) 
\\ [6pt] &=&-\dss\frac{\al}{l} ((\frac{\p}{\p x'^1} - \frac{V}{c^2}\frac{\p}{\p t'})\psi + (-V\frac{\p}{\p x'^1} + \frac{\p}{\p t'})A^1)
\\[6pt] &=& - \dss\frac{\p}{\p x'^1} (\psi'- V A'^1) - \frac{\p A'^1}{\p t'} \ea ,\]

\n
and this gives us the first component of \rf{fields_lor}. A direct computation from \rf{fields_lor} also gives the gauge condition satisfied by these potentials; it is {\sl not}  the Lorenz condition \rf{lgc}, but a formula which appears only in \cite{lor04-} (p. 173):

\be \di'\Ay' + \frac{1}{c^2}\frac{\p \psi'}{\p t'} - \frac{V}{c^2} \frac{\p'A'^1}{\p t'} = 0, \ee{lgctz}

\n
-- another good reason, by the way, {\sl not}  to associate Lorentz's name to the Lorenz condition! Now it is interesting that Lorentz introduces in the same page (\cite{lor04}, p. 814) the retarded potentials for $\rho'_L$ and $\jy'_L$, so one might wonder why the correct Lorenz condition does not follow. The answer is that the derivation of \rf{lgc} uses, crucially, the continuity equation \rf{ce} for the charge -- therefore $\rho'_L$ and $\jy'_L$ do {\sl not}  satisfy \rf{ce}, as was to be pointed out courteously by Poincar\'e (\cite{poi05b}, p. 1506; \cite{poi06}, p. 134). This confirms that Lorentz is not reasoning within a `principle of relativity' perspective, contrary to Minkowski's claim (cf. the second inscription in \S 1; for details \S 4.4, \S 6).

Summing up, it is clear that in \cite{lor04} Lorentz is happy enough to operate at a `low' theoretical level, and that his tinkering with coordinate changes hides nothing particularly profound: he is trying to work out a formalism by which one can best approximate the experimental data. As we shall see (\S 6), Lorentz will explicitly acknowledge the limits of his approach.

\subsection{Poincar\'e, 1905-1906}

In his very wide-ranging article published in 1906 (\cite{poi06}) Poincar\'e chose to start not from the fields but from the potentials, and therefore not from the Maxwell equations, but from the wave equation. Clearly the homogeneous wave equation

\be \Box f  = 0 \ee{we}

\n
where $f$ is a function which can be scalar or vector-valued, is invariant under the conformal special Lorentz transformation \rf{cslt}. In  his section 4 (\cite{poi06}, ``Le groupe de Lorentz") Poincar\'e shows the conformal $l$ factor in \rf{cslt} to be 1 under the assumption that \rf{cslt} make a group, that $l$ be  function of $V$, and that spatial isotropy (in the form of the possibility of a rotation of 180$^o$) holds.\footnote{A different (but unsatisfactory) proof of $l =1$ had been given by Poincar\' e in a letter sent to Lorentz ``during late 1904 to mid-1905'' (it is photographically reproduced in \cite{mi81}, p. 81; by the way, contrary to Miller's statement, this letter does {\sl not} contain ``Poincar\'e's proof that the Lorentz transformations form a group''). The proof is unsatisfactory because it postulates for no good reason that $l$ be of the form $(1-\bt^2)^m$. As a matter of fact, if we reject spatial isotropy, we get a conformal factor of the form \((\dss\frac{1-\bt}{1+\bt})^{r/2}\), as explained in \cite{mmc}.}  Note that, on the other hand, the transformations \rf{voi} do not form a group.\footnote{If one takes as the ``Voigt transformations'' those obtained by composition of one with the inverse of another one among \rf{voi}, one still does not obtain a group (contrary to what is claimed in \cite{eh01}, p. 228), although the {\sl generated} subgroup coincides with the 2-dimensional conformal Lorentz group.}   

If we call $\Ay$ and $\psi$ the vector and the scalar potential, respectively, and define the charge and the current density respectively by $\rho$ and $\jy \equiv \rho \vy$, then the Maxwell equations can be expressed equivalently as \rf{maxp}, subject to \rf{lgc}, the fields being related to the potentials by \rf{fields}. The force density per unit volume created by the electric and magnetic fields is

\be \fy = \rho (\E + \vy\we \B). \ee{force} 

Poincar\'e shows that an elementary (positively or negatively) charged particle (a positive or negative {\sl electron}) with a spherical form in $\phi$ is deformed into an ellipsoidal domain in $\phi'$, and that its volume (i.e. the volume of the interior domain) changes by a factor

\[ \frac{l^3}{\al (1+\frac{V}{c^2}v^1)}. \]

In fact if we consider a ball of radius $R$ moving in $\phi$ with constant velocity $\vy$ and if we call the ball $B_t$ and its boundary $\Si_t$ at the $\phi$-time $t$, we can describe $\Si_t$ in $\phi$ by the equation

\[  (x^1 - v^1 t)^2 + (x^2 - v^2 t)^2 + (x^3 - v^3 t)^2 = R^2. \]

\n
At a fixed time $\phi'$-time $t'$ (Poincar\'e takes $t'= 0$)\footnote{According to some authors the whole of Poincar\'e's treatment in \cite{poi06} falls, at least formally, within the framework of Lorentz's "theorem of corresponding states" (cf. \cite{pb06}); in particular this means that in \cite{poi06} Poincar\'e never changes his reference frame. Although nothing in our analysis depends on this point, we are not convinced. Poincar\'e writes down the equation of the ellipsoid in the primed coordinates (after Lorentz transformation, that is) as a {\sl hypersurface in 4-dimensional space}, that is our \rf{ell}, and then adds: ``Cet ellipsoide se d\'eplace avec un mouvement uniforme; pour $t'= 0$ il se r\'eduit \`a [...]'', which implies that he is interpreting as genuine time coordinates both $t$ and $t'$. So it must be admitted that, while dressing his treatment in a largely Lorentzian fashion, Poincar\'e had still very clear in mind -- and no wonder -- what he had been the first to realize as of 1900: that is, that $t'$ was a genuine physical time.} this corresponds in $\phi'$ to a domain $E_{t'}$ whose boundary $S_{t'}$ is (by application of \rf{cslt_inv}):

\be  \al^2 ((1-Vv^1/c^2) x'^1 - (v^1 - V)t')^2 + (x'^2 -\al Vv^2 x'^1/c^2 - \al v^2 t')^2 +
(x'^3 -\al Vv^3x'^1 /c^2 - \al v^3 t')^2 = l^2 R^2  .\ee{ell}

\n
Surface $S_{t'}$ is homothetic to $\Si_t$, the linear part of the homothety being given by the 3$\times$3 matrix

\[ l \left(\ba{ccc} \al (1- Vv^1/c^2) & 0 & 0 \\
-\al Vv^2 /c^2 & 1 & 0 \\ -\al V v^3/c^2 & 0 & 1\ea\right)^{-1} \]

\n
from which it follows that 

\[ \vo (E_{t'}) = \frac{l^3}{\al (1-Vv^1 /c^2)} \vo (B_t) ,\]

\n
as claimed. This implies, under the assumption of the {\sl invariance of the electric charge} $Q$, supposed to be uniformly distributed within the volume, that the charge density in $\phi'$ is given by:

\be\ba{rcl}  \rho' = \dss \frac{Q}{\vo (E_{t'})} &=& \dss\frac{Q}{l^3 \vo (B_t)}\al (1-Vv^1/c^2) 
\\ [6pt]  &=&  \dss\frac{\al \rho}{l^3} (1- \frac{Vv^1}{c^2})
\\ [6pt]  &=&  \dss\frac{\al}{l^3} (\rho - \frac{V}{c^2} j^1) ,\ea\ee{cd_poi}

\n
which generalizes to the case of an arbitrary charge density by a limiting process. On the other hand the transformation law for the velocity is easily computed directly from \rf{cslt} as

\be \left\{ \ba{rcl} v'^1 &=& \dss\frac{v^1 - V}{1-\frac{V v^1}{c^2}}
\\  v'^2 &=& \dss\frac{v^2}{\al (1-\frac{V v^1}{c^2})}
\\  v'^3 &=& \dss\frac{v^3}{\al (1-\frac{V v^1}{c^2})}, \ea\right. \ee{vel}

\n
whence the transformation law for the current density \( \jy' = \rho'\vy'\) immediately follows:

\be \left\{ \ba{rcl} j'^1 &=& \dss\frac{\al}{l^3} (j^1 - \rho V)
\\  j'^2 &=& \dss\frac{j^2}{l^3}
\\  j'^3 &=& \dss\frac{j^3}{l^3} \ea\right. \ee{cud}

Poincar\'e then proves that with these definitions (but not with those introduced by Lorentz) the continuity equation is invariant. At this point Poincar\'e states the whole of his argument in the following terms (\cite{poi06}, p. 134) :

\begin{quote}
We will define the new potentials, vector and scalar, in such a way as to satisfy the conditions: 

\be \Box' \Ay' = - \mu_0 {\mathbf j'}, \; \Box' \psi' = - \frac{\rho'}{\ep_0}. \ee{maxpr}

Then we will extract from there [``Nous tirerons ensuite de l\`a'']: 

\be \left\{ \ba{rcl} A'^1 &=& \dss\frac{\al}{l} (A^1 - \frac{V}{c^2} \psi)
\\  A'^2 &=& \dss\frac{A^2}{l}
\\  A'^3 &=& \dss\frac{A^3}{l} 
\\ \psi' &=& \dss\frac{\al}{l} (\psi - V A^1).
\ea\right. \ee{pot_p}

\n
These formulas differ from those by Lorentz [that is, from \rf{pot_l}], but the difference is in the last analysis only a matter of definitions.
\end{quote}

\n
Clearly this argument is a little too fast. Previously Poincar\'e had written the formula

\[ \Box'= l^{-2} \Box, \]

\n
and this suggests one conjecture on the path leading to \rf{pot_p}. Let us consider in detail the case of the scalar potential:

\[  \ba{rcl} \Box' \psi' &=& - \dss\frac{\rho'}{\ep_0} = \dss\frac{\al}{l^3} (-\frac{\rho}{\ep_0} + \frac{V}{c^2 \ep_0 \mu_0} \mu_0 j^1) \\  [6pt] &=& \dss\frac{\al}{l^3} \Box (\psi -V A^1 ) \\ [6pt]  &=& \dss\frac{\al}{l} \Box' (\psi -V A^1 ). \ea\] 

\n
It follows that:

\[ \Box' (\psi' -\frac{\al}{l} (\psi -V A^1 )) = 0, \] 

\n
and a similar argument gives:

\[ \Box'(A'^1 - \frac{\al}{l} (A^1 - \frac{V}{c^2} \psi)) = 0, \; 
\Box' (A'^2 - \frac{1}{l} A^2) = 0, \; \Box' (A'^3 - \frac{1}{l} A^3) = 0.\] 

\n
These 4 equations, however, are {\sl not} equivalent to \rf{pot_p}.\footnote{This gap in the hypothetical argument had been pointed out by Schwartz (\cite{sch}, part I, p. 1294, n. 9). Miller (\cite{mi73}, p. 253) does not seem to notice that there is a problem here.} In fact the correct conclusion would be that there are a vector field $\ay$ and a scalar field $b$ such that

\be \left\{ \ba{rcl} A'^1 &=& \dss\frac{\al}{l} (A^1 - \frac{V}{c^2} \psi) + \frac{1}{l}a^1
\\ [6pt] A'^2 &=& \dss\frac{1}{l} (A^2 + a^2)
\\  [6pt] A'^3 &=& \dss\frac{1}{l}(A^3 + a^3) 
\\ [6pt] \psi' &=& \dss\frac{\al}{l} (\psi - V A^1) + \frac{1}{l}b,
\ea\right. \ee{pot_pp}

\n
with $\ay$ and $b$ satisfying

\[ \Box \ay = \0,  \; \Box b = 0, \; \di\ay + \frac{1}{c^2} \frac{\p b}{\p t} = 0. \]

\n
So, according to this reconstruction, what Poincar\'e would have proven is that the potentials transform according to \rf{pot_p} {\sl up to the addition of sourceless potential functions subject to the Lorenz condition}, and the uniqueness of the transformation law would not be completely established.  

\subsubsection{Poincar\'e's argument: a reconstruction}

However, section 5 of \cite{poi06} suggests a different implicit and perfectly valid proof of uniqueness.  In fact there Poincar\'e introduces the retarded potentials \rf{retp} as a way to integrate \rf{maxp} (``It is known that \rf{maxp} can be integrated by the retarded potentials  [...]''). To be sure, we have not to rely on the mention of the retarded potentials (which had been first introduced in print by L. V. Lorenz in 1867) in \cite{poi06} to know that Poincar\'e was perfectly aware of them since at least 15 years, as they are precisely described in a paper of 1891 (\cite{poi91}),\footnote{We owe this reference to \cite{ora}, p. 184.} and discussed also in his lectures on light published in 1892 (\cite{poi92}, pp. 134-9) and in those on electricity and optics published in 1901 (\cite{poi01}, pp. 455-61). Clearly, once the transformation laws for the current and charge densities are assumed (that is, \rf{cd_poi}, \rf{cud}), one obtains e.g. for the scalar potential:

\[ \ba{rcl} \psi' &=&\dss \frac{1}{4\pi\ep_0}\int_{\R^3} [\rho']\frac{\vo'_3}{s'}  
= \dss\frac{1}{4\pi\ep_0}\int_{\R^3}\frac{\al}{l}( [\rho -\frac{V}{c^2} j^1] ) \frac{\vo_3}{s}  
\\ [8pt] &=&  \dss\frac{\al}{l}( \frac{1}{4\pi\ep_0}\int_{\R^3} \frac{[\rho]}{s}\vo_3  -V\frac{\mu_0}{4\pi}\int_{\R^3}\frac{ [j^1] }{s} \vo_3)  
\\ [8pt] &=& \dss\frac{\al}{l} (\psi - VA^1) \ea,\]
 
 \n
The crucial equality, that is

\be \frac{\vo'_3}{s'}  = l^2 \frac{\vo_3}{s},  \ee{retpot}

 \n
 is elementary and can be proven directly as follows. Notice first that \( \vo'_3 = \det J(F) \vo_3\), where  \( F: \R^3 \ra \R^3\) is obtained through the composition
 
 \[ \ry \mapsto (\ry, t_0 -\frac{s}{c}) \stackrel{B}{\mapsto} (\ry', t'_0 -\frac{s'}{c})\mapsto \ry' ,\]
 
 \n
 where $B$ is the conformal special Lorentz transformation \rf{cslt}. Explicitly, 
 
 \[ F(x^1, x^2, x^3) = l (\al (x^1 - V(t_0 -\frac{s}{c})), x^2, x^3) \]
 
 \n
 and therefore
 
 \[ \det J(F) = l^3 \frac{\p x'^1}{\p x^1} = l^3 \al (1 + \frac{V}{c} \frac{x^1 - x^1_0}{s}). \]

\n
On the other hand $s'$ can be expressed in terms of $s$ by noting that:
 
 \[ t'_0 = l\al (t_0 - \frac{V}{c^2}x_0^1) \]
 
 \n
 and therefore
 
 \[ t'_0 - \frac{s'}{c} = l\al (t_0 - \frac{s}{c} -\frac{V}{c^2} x^1) = t'_0 - \frac{l\al s}{c} (1+\frac{V}{c} \frac{x^1 - x^1_0}{s}), \]
 
 \n
 which implies that
 
 \[ s'= l\al s(1 +\frac{V}{c}\frac{x^1 - x^1_0}{s}) = \frac{s}{l^2} \det J(F), \]
 
 \n
so
 
 \[  \frac{\vo'_3}{s'} = (\det J(F) \vo_3) \frac{l^2}{s \det J(F)} = l^2 \frac{\vo_3}{s}, \] 
 
 \n
 as claimed. 
 
 \subsubsection{Commentary}

Poincar\'e may well have thought superfluous to write down this argument, given the uniqueness of the retarded potentials under plausible conditions, but surely this proof was easily within his reach if only he had thought it necessary to provide it. It must be mentioned that Poincar\'e's style of presentation in his technical articles was considered as fraught with gaps in the arguments even by those of his colleagues who most admired him.\footnote{A good piece of evidence is provided by a letter of October 22, 1888, from C. Hermite to G. Mittag-Leffler, where one can read: ``But it must well be admitted that, in this paper as in almost all of his researches, M. Poincar\'e shows well the way and gives some indications, but he leaves a considerable amount to do in order to fill the gaps and to complete his work. Often  Picard has asked him, concerning some very important points in his papers on {\sl Comptes Rendus}, for clarifications and explanations, without being able to obtain more that a statement: `it is this way, it is that way', so that he seems like a seer to whom truths appear in a shining light, but to him only'' (\cite{her}, p. 147).}    

Our reconstruction is strengthened by the circumstance that the approach to the uniqueness of the transformation laws of the fields by the detour of the retarded potentials was explicitly adopted in a short article by an Italian mathematical physicist, Roberto Marcolongo (1862-1943), taking the lead from Poincar\'e \cite{poi06} {\sl in the very same year 1906} (\cite{marc06}), and giving as references for the retarded potentials Poincar\'e's lectures \cite{poi01} and \cite{poi92}. Although Marcolongo does not prove \rf{retpot} directly, his paper can be described as clarifying Poincar\'e's  implicit argument, the avowed scope of the paper being to present a means for finding solutions of the Maxwell equations ``very easily and elegantly'', by using as a basic ingredient the retarded potentials. The last section of \cite{marc06} deals a little hurriedly with the transformation laws under the Lorentz group: it is worth noting, however,  that Marcolongo, who does not seem to have known Einstein's paper at the time, linked the 4-dimensional formalism with the transformation laws for the potentials more explicitly than Poincar\'e had done, and well in advance with respect to Minkowski.\footnote{Notice that the possibility of obtaining the retarded potentials from a single 4-dimensional potential (i.e. a function of the space and time variables) had been proved, in a more roundabout fashion, {\sl before the rise of special relativity} by Gustav Herglotz in 1904 (\cite{her04}; cf. \cite{pau}, p. 90; \cite{som}, pp. 245-9).}    
  
Our analysis gives an important result: Poincar\'e's argument for the transformation laws of the fields may be considered as essentially complete, if  we grant him the assumption that the retarded potentials are the correct solutions of the wave equations, thus ruling out in one blow: 1) sourceless potentials, 2) advanced potentials. Both exclusions are consistent with Poincar\'e's approach, and the second one is explicitly stated, on grounds of causality, in the final section of the paper.\footnote{``We will endeavour to make it so that $t$ will always be negative; in fact, if one conceives that the effect of gravitation takes some time to propagate, it would be more difficult to understand how this effect could depend on the position {\sl not yet arrived at} of the attracting body'' (\cite{poi06}, p. 167).} It is true that, as Walther Ritz in 1908 remarked, the primacy given to potentials with respect to the system of the Maxwell equations lands us into another theory with respect to Maxwell electromagnetism (\cite{ritz}, p. 171):

\begin{quote}
[\ldots] Maxwell would see an essential advantage of his theory precisely in the fact that it dispenses with any consideration of elementary actions or of the origin of the field, and it deals only with the immediate neighborhood of the given point. We see that this is not the case, and that to eliminate physically impossible solutions, one must adopt a priori the retarded potentials formulas, which distinguish the elementary actions as in the classical theories, and verify that they satisfy the equations, i.e. that they can completely replace the equations, while the reverse is not true.
\end{quote}

\n
However, as Lorentz pointed out in his 1906 lectures at the Columbia university, ``[w]e need not however speak of other solutions, if we assume that an electromagnetic field in the ether is never produced by any other causes than the presence and motion of electrons''.\footnote{\cite{lor09}, p. 20. As remembered by O'Rahilly in his useful and abundantly referenced account of ``propagated potentials'' in chapter VI of  his treatise: ``It is not generally realized nowadays that there was considerable opposition to the introduction of potential'' (\cite{ora}, p. 182). A recent attempt at basing electrodynamics on the potentials, and a response to some of the standard `textbook' objections, with several references, is contained in \cite{rou04}.} 

We shall see later that a more serious stumbling block occurs in Einstein's deduction of the transformation laws for the fields.   

As to Poincar\'e's comment that his difference from Lorentz ``is in the last analysis only a matter of definitions'', this is surely a gesture of leniency towards his illustrious colleague rather than a claim to be taken literally, unless we assume that for Poincar\'e charge continuity itself  was ``a matter of definitions''.  

\subsubsection{The transformation laws for the fields and the field invariants}

Granted  \rf{pot_p}, the transformation law for the fields are then obtained by simply adapting \rf{fields} to the primed system and taking account of \rf{par}; thus one obtains

\be \left\{\ba{rcl} E'^1 &=& l (V)^{-2} E^1  \\ [4pt]              
E'^2 &=&  l (V)^{-2} \al (E^2 -V B^3)  \\ [4pt]
E'^3 &=& l (V)^{-2}\al (E^3+V B^2)  \ea \right. , 
\left\{\ba{rcl}   B'^1 &=& l (V)^{-2} B^1 \\ [4pt] B'^2 &=&  l (V)^{-2} \al (B^2 + \frac{V}{c^2} E^3)  
\\ [4pt]  B'^3 &=& l (V)^{-2} \al (B^3 -\frac{V}{c^2} E^2), \ea\right.  
\ee{ft3}  

\n
which in turn can be used to derive the transformation law for the standard force density, i.e. the force density {\sl per unit volume}. For instance:

\[ \ba{rcl} f'^1 &=& \rho' (E'^1 + v'^2 B'^3 - v'^3 B'^2) \\ [4pt]
&=& \dss\frac{\al\rho}{l^5} (1-\frac{Vv^1}{c^2}) (E^1 + \frac{1}{1- Vv^1 /c^2} (B^2 - \frac{V}{c^2}E^2) - \frac{v^3}{1-Vv^1 /c^2} (B^2 + \frac{V}{c^2} E^3)) \\ [4pt]
&=& \dss\frac{\al}{l^5} (\rho (\E + \vy\we\B)^1 -\frac{\rho V}{c^2} (\E\cdot \vy))\ea, \]

\n
and by taking into account $\rho \E\cdot\vy = \fy \cdot \vy$ we have, by similar computations, the system:

\be \left\{ \ba{rcl} f'^1 &=& \dss\frac{\al}{l^5} (f^1 - \frac{V}{c^2} \fy\cdot\vy) \\ [4pt]
 f'^2 &=& \dss\frac{1}{l^5} f^2 \\ [4pt] 
f'^3 &=& \dss\frac{1}{l^5} f^3 \ea\right.  ,\ee{fot}

\n 
while the transformation equations for the force densities {\sl per unit charge} are:
  
\be \left\{ \ba{rcl} F'^1 &=& \dss\frac{\al}{l^5} \frac{\rho}{\rho'}(F^1 - \frac{V}{c^2} \Fy\cdot\vy) \\ [4pt]
 F'^2 &=& \dss\frac{1}{l^5} \frac{\rho}{\rho'}F^2 \\ [4pt] 
F'^3 &=& \dss\frac{1}{l^5} \frac{\rho}{\rho'}F^3 \ea\right.  .\ee{fotc}  
  
From \rf{ft3} Poincar\'e proves that the action integral

\[ J = \frac{1}{2}\int (|\E|^2 - |\B|^2)\vo_4 \]

\n
is a conformal Lorentz invariant. This follows from \(\vo'_4 = l^4\vo_4\) and the identity

\[ l^4 ((|\E'|^2 - |\B'|^2) = (|\E|^2 - |\B|^2).\]

\n
Poincar\'e also writes the other invariance identity:

\[ l^4 \E'\cdot\B'= \E\cdot\B. \]

\n
As we have said, in \S 4 of \cite{poi06} Poincar\'e proves that $l =1$, so that, in particular:

\be \left\{\ba{rcl} E'^1 &=& E^1  \\ [4pt]              
E'^2 &=&  \al (E^2 -V B^3)  \\ [4pt]
E'^3 &=& \al (E^3+V B^2)  \ea \right. , 
\left\{\ba{rcl}   B'^1 &=& B^1 \\ [4pt] B'^2 &=&  \al (B^2 + \dss\frac{V}{c^2} E^3)  
\\ [6pt]  B'^3 &=& \al (B^3 -\dss\frac{V}{c^2} E^2). \ea\right.  
\ee{ft3bis}  
  
Poincar\'e is particularly concerned with the physical possibility of a stable ``electron'', that is, an elementary (positively or negatively) charged particle. From \rf{fot} it follows that the condition of equilibrium $\fy = \0$ is invariant under the conformal Lorentz transformation. However, in an admissible system in which the electron is at rest, this condition implies $\E = \0$ and therefore, by the Gauss law for the electric field, $\rho = 0$. In other words, if all forces acting on the electron were of electromagnetic origin (i.e. were included in the expression for the Lorentz force law), then {\sl a static charged particle would be impossible}. Conversely, if we assume that electrons exist, then other constraints must be involved in their dynamical balance. About a third of Poincar\'e's paper is devoted to analysing the exact nature of these constraints, and their compatibility with the principle of relativity. He shows that a potential has to be introduced to account for the deformation that an electron must undergo according to Lorentz's theory as opposed to Max Abraham's and Paul Langevin's alternative proposals, and that this potential is proportional to the volume of the electron (which of course, according to Lorentz, depends on velocity, while Langevin assumed it to be constant). This part of the article is a sophisticated contribution to what would be classified today as particle physics.\footnote{An outline of the historical background is provided in \cite{mar05}.} Nothing of comparable depth on this topic, as judged by the standards of the time,  can be found in Einstein's article \cite{ei05}.\footnote{We shall expand on this comment at the end of \S 4.3.} 

\subsubsection{Poincar\'e and 4-dimensional space}

In the last section of \cite{poi06} -- the section devoted to gravitation -- Poincar\'e introduces explicitly the concept of  $(x^1, x^2, x^3, x^4 \equiv ict)$ being ``coordinates'' in ``the space with 4 dimensions'' and goes on by making a memorable statement from which, as we shall see, Minkowski will profit: ``We see that the Lorentz transformation is nothing but a rotation in this space around the origin, viewed as fixed'' (\cite{poi06}, p. 168). 

In modern terminology, the use of imaginary time (that is, $x^4 = ict$) amounts to representing the proper orthochronous Lorentz group $\LCU^+$ as a subgroup of the complex orthogonal  group $O(4, \C)$. In fact, by identifying  \( \R^4 \cong \R^3 \times i\R \), every matrix $\La \in \LCU^+$ is represented as

\[ \La^\si = \Si \La \Si^{-1}, \; \mbox{where} \; \Si = \left( \ba{cc} I_3 & 0 \\ 0 & ic\ea\right). \]   

\n
Formally, we have a group monomorphism:\footnote{Clearly $\det\La^\si = \det \La = 1$, so in fact $\La^\si\in SO(4,\C)$.} 

\[\ba{rcl}  \LCU^+ &\ra & SO (4, \C)
\\ [8pt] \La = \left( \ba{cc} A & -A\Vy  \\ -\frac{\al}{c^2} \Vy^T & \al\ea\right) &\mapsto&  
\La^\si  = \left( \ba{cc} A & -\frac{i}{c} A\Vy  \\ -\frac{i\al}{c} \Vy^T & \al\ea\right). \ea\] 

Poincar\'e does not refer to the concept of a tensor and does not use the absolute calculus which Ricci-Curbastro, following a remark by Christoffel, had introduced since 1892 and developed together with his pupil, Levi-Civita (\cite{rclc}). This is particularly intriguing, since in the preface of their joint work the two Italian mathematicians had stressed the value of the tensor formalism citing Poincar\'e himself to the effect that ``a good notation has the same philosophical importance as a good classification in the natural sciences''. However, from Poincar\'e's own viewpoint, there was only a linguistic divide between what he did in his paper and the formal recognition that the electric and magnetic fields had been proven to be `parts' of a double tensor in 4-dimensional space. Notice that in contrast to Lorentz, Poincar\'e did not use the {\sl vector} formalism either: he always dealt with Cartesian components, in agreement with the French common usage.\footnote{``While the first German textbook on electromagnetism to employ vector notation systematically dates from 1894 (F\"oppl, 1894), the first comparable textbook in French was published two decades later by Jean-Baptiste Pomey $(1861-1943)$, instructor of theoretical electricity at the {\sl \'Ecole sup\'erieure des Postes et T\'el\'egraphes} in Paris (Pomey, $1914-1931$, vol. 1)'' (\cite{wal06}, p. 200n19).} For instance, here is how Poincar\'e introduces the basic vector quantities in his paper (\cite{poi06}, p. 132):

\begin{quote}
[...] calling $f,g,h$ the electric displacement, $\alpha, \bt, \g$  the magnetic force, $F, G, H$ the potential vector, [...] $\xi, \eta, \zeta$ the velocity of the electron, $u, v,w$ the current [...]
\end{quote}

\n
Moreover, Poincar\'e often writes a vector equation by singling out the first component only: for instance he writes the inhomogeneous wave equation for the potential vector (i.e. the first of our \rf{maxp}) simply as $\Box F = \rho\xi$ ({\sl ibid.}). In dealing with authors such as Lorentz or Abraham, who currently used the vector calculus formalism, Poincar\'e does not bother to establish a `dictionary' from their notation to his, clearly being confident that his readers would not have found any difficulty in making the needed translation by themselves. In his {\sl Treatise} J. C. Maxwell had written:

\begin{quote}
But for many purposes of physical reasoning, as distinguished from calculation, it is desiderable to avoid explicitly introducing the Cartesian coordinates, and to fix the mind at once on a point of space instead of its three coordinates, and on the magnitude and direction  of a force instead of its three components. [...] As the methods of Des Cartes are still the most familiar to students of science, and as they are really the most useful for purposes of calculation, we shall express all our results in the Cartesian form.\footnote{\cite{max}, p. 9; O. Heaviside adopted systematically the vector formalism, and he emphasized in the first volume of his {\sl Electromagnetic Theory} (London, 1893) that: ``My system, so far from being inimical to the cartesian system of mathematics, is its very essence'' (cit. in \cite{crowe}, p. 173).} 
\end{quote}

\n
Poincar\'e's approach to the vector formalism seems to have been similar to Maxwell's; a consequence of this is lack of interest in making explicit the conceptual nature of what are, for all purposes, 4-dimensional tensor equations.\footnote{In the same spirit Schwartz wrote: ``Had Poincar\'e adopted the ordinary vector calculus that was already in use by theoretical physicists -- for example, Lorentz and Abraham -- for some time, he would have in all likelihood introduced explicitly in the present connection the convenient four-dimensional vector calculus'' (\cite{sch}, III, p. 1287fn.7).} On the other hand, what is missing in his treatment, with respect to what Minkowski will do, is the use of 4-dimensional differential operators to reformulate the Maxwell equations. 
 
\subsection{Einstein, 1905}

As is well known, in his \cite{ei05} Einstein started from two postulates: the principle of relativity and the constancy of the velocity of light in a given, ``stationary'' coordinate system. However, ``Maxwell's theory'' is mentioned already in the introductory section (although the ``Maxwell-Hertz equations'' are  to be found in a later section of the paper, i.e. \S 6):

\begin{quote}
These two postulates suffice for the attainment of a simple and consistent theory of the electrodynamicas of moving bodies based on Maxwell's theory for stationary bodies. 
\end{quote}

\n
This statement is strange, since the constancy law is a consequence of Maxwell's theory taken together with the principle of relativity, so there appears to be some redundancy.\footnote{On the other hand it is obvious that from the two postulates the Maxwell equations cannot be derived.} Einstein was aware that Maxwell's (or Lorentz's) electromagnetism could not be considered as a sufficiently solid ground on which to build the whole of physics. In fact his `quantum' paper of the same year (the paper he called ``very revolutionary'') introduced a sharp departure from classical electromagnetism, so in Einstein's thought light's behaviour had a foundational primacy with respect to electromagnetism (\cite{abi05a}, pp. 81-5; cf. \cite{pai82}, pp. 139, 147). 

From his two postulates Einstein derived \rf{cslt}. Considering the Maxwell equations in empty space (Einstein assumes the Gauss laws tacitly; he will write down the Gauss law for the electric field only in his section 9):

\be \left\{\ba{rcl} \cu \E &=& -\dss\frac{\p\B}{\p t} \\ [8pt] \cu\B &=& \dss\frac{1}{c^2}\frac{\p\E}{\p t} \\ [8pt] \di\E &=& 0 \\ [8pt] \di\B &=& 0 \ea\right. , \ee{max_0}

Einstein deduces that 

\be \left\{\ba{rcl} \cu' \td{\E} &=& -\dss\frac{\p \td{\B}}{\p t'} \\ [8pt] \cu'\td{\B} &=& \dss\frac{1}{c^2}\frac{\p\td{\E}}{\p t'} \\ [8pt] \di'\td{\E} &=& 0 \\ [8pt] \di'\td{\B} &=& 0 \ea\right. ,\ee{max_td}

\n
where the operators $\cu'$ and $\di'$ are meant with respect to $\phi'$ and $\td{E}$ and $\td{\B}$ are defined by

\be \left\{\ba{rcl}\td{E}^1 &=& \psi (V) E^1  \\ [4pt]              
\td{E}^2 &=&  \psi (V)\al (E^2 -V B^3)  \\ [4pt]
\td{E}^3 &=& \psi (V)\al (E^3+V B^2)  \ea \right. , 
\left\{\ba{rcl} \td{B}^1 &=& \psi (V) B^1 \\ [4pt] \td{B}^2 &=&  \psi (V) \al (B^2 + \frac{V}{c^2} E^3)  
\\ [4pt]  \td{B}^3 &=& \psi (V)\al (B^3 -\frac{V}{c^2} E^2), \ea\right.  
\ee{ft}  

\n
where $\psi (V)$ is a function of $V\in ]-c,c[$. Einstein states, with reference to \rf{max_td} and to the `primed' version of  \rf{max_0} (which is a direct consequence of the relativity principle):

\begin{quote}
Evidently the two systems of equations found for system [$\phi'$] must express exactly the same thing, since both systems of equations are equivalent to the Maxwell-Hertz equations for system [$\phi$].  Since, further, the equations of the two systems agree, with the exception of the symbols for the vectors, it follows that the functions occurring in the systems of equations at corresponding places must agree, with the exception of a factor $\psi (V)$, which is common for all functions of the one system of equations, and is independent of  [$x'^1$, $x'^2$, $x'^3$ and $t'$] but depends upon $[V]$. [\cite{ei05}, p. 908; \cite{ei}, p. 53]
\end{quote}

\n 
Thus Einstein infers that $\td{\E} = \E'$ and $\td{\B} = \B'$, that is \rf{ft3bis}, except for the factor $\psi (V)$ which he will 
subsequently show to be equal to 1 (see below).

This argument is defective, although most commentators seem to be happy with it.\footnote{For instance, in his line-by-line analysis of \cite{ei05}, Miller has nothing to say on this point (\cite{mi81}, pp. 287-8); Torretti (\cite{tor83}, p. 109) neglects the uniqueness issue; Zahar (\cite{za89}, pp. 113-6) sees that there is a problem, but he thinks that he can solve it by using an argument which just mirrors in more formal terms Einstein's own argument.} No doubt, \rf{ft} is a {\sl sufficient} condition for the covariance of \rf{max} under \rf{cslt}: but in this passage Einstein is implying its {\sl necessity}.  Vanishing electric and magnetic fields in $\phi$ might for instance transform into {\sl constant} fields in $\phi'$, which obviously would preserve \rf{max}; in fact it is well-known that \rf{max} admits also non-constant solutions.\footnote{Even the first possibility (zero fields in $\phi$ transforming into nonzero constant fields in $\phi'$) cannot be immediately dismissed: an `inertial' equivalence principle might be valid for electromagnetism; it is only by explicitly taking into account the expression for the electromagnetic force (the Lorentz force law) that we can rule out this possibility.}  

One might recur to a classic uniqueness theorem on the Cauchy problem for Maxwell equations in empty space (\cite{ch}, pp. 647-8), which in modern relativistic terms (\cite{sw77}, p. 124) can be expressed as follows:

\begin{thm} Let $\E$ and $\B$ be solutions of \rf{max_0}, and let $J^- (\0, k)$ be the causal past of an event $(\0, k)$ with $k>0$ in $\R^4$ viewed as the standard Minkowski space-time. If on the domain $J^- (\0, k) \cap \{ t= 0\}$ both $\E$ and $\B$ vanish, then they vanish also on  $J^- (\0, k) \cap \{ t > 0\}$.\end{thm}
 
\n
This theorem implies that if we assume that \rf{ft} holds on $J^- (\0, k) \cap \{ t= 0\}$, then the only solution of \rf{max} compatible with vanishing initial data on the corresponding 3-surface $J^- (\0, \al k) \cap \{ t' = 0\}$ is the zero one. Obviously {\sl any} linear combination of zero fields in $\phi$ would make a suitable choice for the corresponding fields in $\phi'$... One might think of {\sl nonvanishing} initial data, but in order to apply the previous theorem we would need to know that the correct transformation formula for the fields on the initial surface is \rf{ft}, which begs the question. 

Moreover, in case one accepts \rf{ft} at the initial data level, it is unclear why one should take the factor $\psi (V)$ not to be equal to 1 {\sl from the start}. In fact in this case Einstein's argument to the effect that $\psi (V) \equiv 1$ would be a clumsy repetition of the argument (based on spatial isotropy) by which he established in a previous section that the conformal factor $l$ in the Lorentz special transformation must be identically 1. Here is what Einstein writes (\cite{ei05}, p. 59; \cite{ei}, p. 53, italics added):

\begin{quote}
If we now form the reciprocal of [\rf{ft}] by solving the equations just obtained, and secondly by applying the equations to the inverse transformation (from [$\phi'$]  to [$\phi$]), which is characterized by the velocity $-V$, it follows, when we consider that the two systems of equations thus obtained must be identical, that $\psi (V) \psi (-V) =1$. {\sl Further for reasons of symmetry} [{\sc footnote}] $\psi (V) = \psi(-V)$ and therefore $\psi (V) =1$ [...]
\end{quote}

\n
The footnote reads, in full:

\begin{quote}
If, for example, [$E^1 = E^2 = E^3 = 0, B^1 = B^2 = 0$] and [$B^3 \neq 0$], then from reasons of symmetry it is clear that when $V$ changes sign without changing its numerical value, [$B'^2$] must also change sign without changing its numerical value.
\end{quote} 

Equations \rf{ft}, after substituting  $\td{\E} \leadsto \E'$ and $\td{\B} \leadsto \B'$, lead to 

\be \left\{\ba{rcl} E^1 &=& \psi (V)^{-1} E'^1  \\ [4pt]              
E^2 &=&  \psi (V)^{-1} \al (E'^2 +V B'^3)  \\ [4pt]
E^3 &=& \psi (V)^{-1} \al (E'^3 - V B'^2)  \ea \right. , 
\left\{\ba{rcl}   B^1 &=& \psi (V)^{-1} B'^1 \\ [4pt] B^2 &=&  \psi (V) \al (B'^2 - \frac{V}{c^2} E'^3)  
\\ [4pt]  B^3 &=& \psi (V)^{-1}\al (B'^3 +\frac{V}{c^2} E'^2), \ea\right.  
\ee{ft2}  

\n
and therefore $\psi (V)\psi (-V) = 1$. At this point the footnote we have cited comes in, introducing ``reasons of symmetry'' in the sense of what may be called the {\sl principle of reciprocity of effects}, which is, in fact, equivalent to spatial isotropy (and  must be kept distinct from what is commonly known as the `reciprocity principle'),\footnote{See \cite{mmc}, pp. 1393-5, for an explanation.} thus giving $\psi (-V) = \psi (V)$ and therefore $\psi (V) =1$. Einstein does not refer to the introductory section of his paper, but it is clear that it is this kind of symmetry which he has in mind.\footnote{For an analysis of Einstein's `conductor and magnet' argument see \cite{bm}.}

The very fact that Einstein needs to prove separately that $l=1$ and $\psi=1$ shows that for him the Lorentz transformation does not apply directly to the pair $(\E, \B)$: in other words he is not aware that this pair is in some sense a generalized vectorial entity. Actually, in \cite{ei05} the concept of space-time as a geometric space in its own right does not appear, even implicitly.

In section 9 Einstein uses the Gauss law for the electric field to derive the transformation law for the charge density; in fact, using \rf{par} with $l=1$ we find 

\be \ba{rcl} \rho' &=& \ep_0 \di'\E'
\\ [8pt] &=& \ep_0 \al \dss(\frac{\p E^1}{\p x^1} + \frac{V}{c^2} \frac{\p E^1}{\p t} + \frac{\p}{\p x^2}(E^2 - VB^3)
+ \frac{\p}{\p x^3} (E^3 + VB^2)) 
\\ [8pt]  &=& \ep_0 \al (\di\E +V(\cu\B -\dss\frac{1}{c^2}\frac{\p\E}{\p t})^1)
\\ [8pt] &=& \ep_0 \dss\al (\frac{\rho}{\ep_0} -V\mu_0 j^1) = \al (\rho -\frac{V}{c^2} j^1) 
\ea \ee{cd_ei}    
 
 \n
 which is the same formula obtained by Poincar\'e, though by a different route.
 
 \subsubsection{Einstein's electron}
 
 It is interesting that the very word `electron' is used by Einstein to mean just ``an electrically charged particle'', with the warning that ``a ponderable material point can be made into an electron (in our sense of the word) by the addition of an electric charge, {\sl no matter how small}'' (\cite{ei05}, \S 10, italics in the original). The whole issue of the shape and stability of the electron as a finite body, and the way motion modifies it -- an issue which was at the centre of the concerns of the main scientists in the field, for both theoretical and experimental reasons -- is entirely neglected, although Einstein touched on the contemporary debate with his derivation of the transverse and longitudinal masses of the electron.\footnote{His formula for the transverse mass (\cite{ei05}, p. 919) is incorrect.} 

When, in 1906, Einstein suggests a new method to sort out experimentally the different predictions of the contemporary theories of the electron, he lists them as the ``theory of Bucherer'', the ``theory of Abraham'' and the ``theory of Lorentz and Einstein'' (\cite{ei06}), thus implicitly accepting Lorentz's  speculations on the electron as naturally integrable in his own theory of relativity.\footnote{Cf. \cite{sta}, p. 274. Pais (\cite{pai82}, pp. 155, 159) presents Einstein's omission of a treatment of the electron as an elementary particle in 1905 as if in a way his mass-energy equivalence equation had obviated the problems that obsessed some of his eminent colleagues; however he adds that the problem is still far from solution or even from a proper formulation: ``The investigations of the self-energy problem by men like Abraham, Lorentz, and Poincar\'e have long since ceased to be relevant. All that has remained from these early times is that we still do not understand the problem. [...] Special relativity killed the classical dream of using the energy-momentum-velocity relations of a particle as a means of probing the dynamic origins of its mass. The relations are purely kinematic. [...] But we still do not know what causes the electron to weigh''. For more on the development of the problem see \cite{fey}, chapters 27-28 of vol. II; \cite{mi73}, pp. 303-19; \cite{roh90}, chapter 2.}  Moreover, at a time when relativity had been developed to its mature form, Einstein tackled the stability problem with an approach similar to Poincar\'e's (but with no reference to him) in a paper of 1919 (\cite{ei19}, translated in \cite{ei}), where he considers the possibility that the key to the ``equilibrium of the electricity constituting the electron'' may lie in the ``gravitational forces''.\footnote{``Therefore by equation (1) [i.e. the field equation of general relativity], we cannot arrive at a theory of the electron by restricting ourselves to the electromagnetic components of the Maxwell-Lorentz theory, {\sl as has long been known}. Thus if we hold to (1) we are driven on to the path of Mie's theory'' (\cite{ei}, p. 193; italics added).}  
 
\subsection{Minkowski, 1907-1908} 

In the paper ``The fundamental equations for the electromagnetic processes in moving bodies'' (\cite{min08a}), which is a development and a fuller treatment of results presented in \cite{min15}, Minkowski systematically uses and develops the 4-dimensional formalism, with `imaginary' time, introduced by Poincar\'e, but, although Minkowski refers twice to \cite{poi06}, he fails to acknowledge properly his debt to it, as we shall emphasize in the last section; in particular he claims originality for the use of the imaginary time variable (\cite{min08a}, p. 56)! 

In the following, adopting the notation introduced at the end of \S 6,  we assume that $x'= \La^\si x$ (while in Minkowski's paper it is the other way around).\footnote{With minor changes, we use for the quotations the Wikipedia translation \cite{min08a_wiki}.}  

In \S 2-\cite{min08a}, Minkowski splits the Maxwell equations into two subsystems, the `inhomogeneous' and the `homogeneous' one, which he writes as follows:

\be \left\{\ba{rcl} \cu\Hy -\dss\frac{\p\Dy}{\p t} &=&  \jy \\ [8pt]  \di \Dy &=& \dss \rho . \ea\right. .\ee{max1}

\be \left\{\ba{rcl} \cu \E +\dss\frac{\p\B}{\p t} &=& 0 \\ [8pt]   \di \B &=& 0, \ea\right. \ee{max2}

By using the imaginary coordinate $x^4$ they can be re-written as, respectively:

\be \left\{\ba{rcl} \cu\Hy -\dss\frac{\p (ic\Dy)}{\p x^4} &=& \jy \\ [8pt]  \di (ic\Dy) &=& ic\rho  . \ea\right. \ee{max1a}

\be \left\{\ba{rcl} \cu (i\E/c) -\dss\frac{\p\B}{\p x^4} &=& 0 \\ [8pt]  \di \B &=& 0  , \ea\right. \ee{max2a}

\n
In this reformulation the left-hand sides of the two subsystems are perfectly similar from a mathematical point of view. Let us write down the components:

\be \left\{\ba{rcl}  
\dss\frac{\p H^3}{\p x^2} - \frac{\p H^2}{\p x^3} -  \frac{\p (ic D^1)}{\p x^4} &=&  j^1 
\\ [8pt]  \dss  - \frac{\p H^3}{\p x^1} + \frac{\p H^1}{\p x^3} -  \frac{\p (ic D^2)}{\p x^4} &=&  j^2
\\ [8pt]  \dss\frac{\p H^2}{\p x^1} - \frac{\p H^1}{\p x^2} -  \frac{\p (ic D^3)}{\p x^4} &=&  j^3
\\ [8pt] \dss\frac{\p (ic D^1)}{\p x^1} + \frac{\p (ic D^2)}{\p x^2} + \frac{\p (ic D^3)}{\p x^3} &=& ic\rho
\ea\right. .\ee{max1b}

\be \left\{\ba{rcl}  \dss\frac{\p (i E^3 /c)}{\p x^2} - \frac{\p (iE^2/c)}{\p x^3} -  \frac{\p B^1}{\p x^4} &=& 0  
\\ [8pt]  \dss  - \frac{\p (i E^3/c)}{\p x^1} + \frac{\p (iE^1/c)}{\p x^3}-  \frac{\p B^2}{\p x^4} &=& 0
\\ [8pt]  \dss\frac{\p (iE^2/c)}{\p x^1} - \frac{\p (i E^1/c)}{\p x^2} -  \frac{\p B^3}{\p x^4} &=& 0 
\\ [8pt] \dss\frac{\p B^1}{\p x^1} + \frac{\p B^2}{\p x^2} + \frac{\p B^3}{\p x^3} &=& 0 
\ea\right. \ee{max2b}

\n
Minkowski names \S 3-\cite{min08a} ``The relativity theorem by Lorentz" ({\sl Das Theorem der Relativit\"at von Lorentz}).  What is this theorem? The explanation is to be found in the introduction to the paper: 

\begin{quote}
The covariance of these fundamental equations under the Lorentz transformations is a purely mathematical fact; I will call this the Theorem of Relativity; this theorem rests essentially on the form of the differential equations for the propagation of waves with the velocity of light. [...]

H. A. Lorentz has found out the ``Relativity theorem" and has created the Relativity-postulate as a hypothesis that electrons and matter suffer contractions in consequence of their motion according to a certain law. [\cite{min08a}, pp. 54, 55]
\end{quote}

\n
From a logical point of view, to say that the ``covariance of these fundamental equations under Lorentz transformations is a purely mathematical fact'', that is, not based on any physical assumptions, is incorrect, since it neglects the necessary and physically substantive adjustment that the transformation laws for the charge and current densities must undergo (cf. \cite{bm}, \cite{bm+}).  But the historical component of this claim is close to astonishing: as we know, Lorentz neither found out a ``relativity theorem'' in this sense, nor even stated a postulate of relativity. Both attributions should be re-directed to Poincar\'e. Moreover, Minkowski's proof (and his whole approach) systematically exploits the interpretation of the Lorentz transformations as rotations in 4-dimensional space, which is again something for which he is indebted to Poincar\'e. To see how, we have to make a detour. 

Minkowski does not refer to the general formalism introduced by Ricci-Curbastro (\cite{rclc}),\footnote{Notice that the use of tensors had been recommended to theoretical physicists by Voigt at least since 1898 (\cite{jmc}, II, p. 274).}  but in \S 5-\cite{min08a} he defines in a separate way the 4-dimensional objects he needs, which he calls {\sl vectors of kind I} and {\sl vectors of kind II}, according to how they trasform under a Lorentz transformation. 

The vectors of kind I are just those ``systems of 4 quantities" which vary under a Lorentz transformation as coordinates do -- that is, they are free (co-)vectors in a 4-dimensional affine space. The vectors of kind II (previously called ``tractors'' in \cite{min15}, p. 933) are, at first, defined as 6-tuples $(f_{23}, f_{31}, f_{12}, f_{14}, f_{24}, f_{34})$ which vary under a Lorentz transformation so as to leave the following expression invariant: 

\be \ba{rcl} f_{23} (x^2 u^3 - x^3 u^2) &+& f_{31} (x^3 u^1 - x^1 u^3) + f_{12} (x^1 u^2 - x^2 u^1)
\\ [4pt] + f_{14} (x^1 u^4 - x^4 u^1) &+& f_{24} (x^2 u^4 - x^4 u^2) + f_{34} (x^3 u^4 - x^4 u^3) \ea \ee{min}  

\n
in passing from the unprimed to the primed coordinate system, for every choice of $x^i$ and $u^i$ among the vectors of kind I. Here and occasionally in the following we distinguish the covariant from the contravariant indices, while Minkowski, thanks to the `Euclidean' translation of the Lorentzian structure, identifies them.   

Now the invariance of \rf{min}, once the nature of the 6-tuples as double covariant antisymmetric tensors (or, as they are today also called, 2-forms) is recognized, is just a special case, as is easy to verify, of the standard rule to obtain the transformation law for {\sl any} tensor:  let $T_{i_1 \ldots i_p}^{j_1 \ldots j_q}$ be the components of a tensor of type  $(q,p)$ in the unprimed system, then the components  ${T'}_{i_1 \ldots i_p}^{j_1 \ldots j_q}$ in the primed system are uniquely defined by the condition 

 \[ T_{i_1 \ldots i_p}^{j_1 \ldots j_q} u_1^{i_1}\ldots u_p^{i_p}\te^1_{j_1} \ldots \te^q_{j_q} = 
{T'}_{i_1 \ldots i_p}^{j_1 \ldots j_q}  {u'}_1^{i_1}\ldots {u'}_p^{i_p}{\te'}^1_{j_1} \ldots {\te'}^q _{j_q} \]

\n
where  $u_1,\ldots, u_p$ are vectors and $\te^1, \ldots, \te^q$ are co-vectors.  As a vector of kind II, the transformation law of $f$ is 

\be f'= \La^\si f (\La^\si)^T \ee{IIk}

\n
which gives for $\E$ and $\B$ the correct relativistic transformation laws once they are identified with suitable sets of components of $f$, as we shall see in a moment. 

The interpretation of the vectors of kind II as ``alternating'' (i.e. skew-symmetric) $4\times 4$ matrices is anticipated in \S 3-\cite{min08a} and formalized in \S 11-\cite{min08a}, where the concept of the ``dual'' matrix is also introduced:

\be f^{\ast} = (f^{\ast}_{ij}) \; \mbox{with}\; f^{\ast}_{ij}= f_{kl}, \; (ijkl)\; \mbox{even permutation of} \; (1, 2, 3, 4) . \ee{dual}

\n
Thus

\[ f^\ast = \left(\ba{cccc} 0  & f_{34} & f_{42} & f_{23} \\  f_{43} & 0 & f_{14} & f_{31} 
\\ f_{24} & f_{41}& 0 & f_{12}  \\f_{32} & f_{13} & f_{21} & 0 \ea\right). \]

\n
The reasoning behind the introduction of these algebraic entities\footnote{Whittaker indicated a plausible anticipation of the vectors of kind II in the 
Pl\"ucker-Cayley coordinates of 1868-9 (\cite{whi53}, vol. II,  pp. 34-5). } can be reconstructed as follows from \rf{max1b} and \rf{max2b}. In \S 12-\cite{min08a} of \cite{min08a} a 4-dimensional divergence operator is introduced in the natural way, Minkowski denoting it by `lor':

\[ \lo = (\p_1, \p_2, \p_3, \p_4) = \p, \] 

\n
where

\[ \p_\al = \frac{\p}{\p x^\al} \, (\al =1,2,3), \;\; \p_4 = \frac{\p}{\p (ict)} = -\frac{i}{c}\frac{\p}{\p t} .\]

\n 
The way  $\lo$ acts is simply by operatorial matrix product on vectors (or, more exactly, vector fields) of both kinds I and II:

\[ \lo S : = \p S. \]

Notice that

\[ \p =\p' \La^\si \]

\n
and therefore if $S$ transforms according to \rf{IIk} we have

\[ \p S = (\p'\La^\si)(\La^\si)^T S' \La^\si = \p' S'\La^{\si}, \]

\n
which is how a co-vector should vary under a coordinate change. Now $\lo f$ for a vector $f$ of kind II is equivalent to  the 4-tuple

 \be \left(\ba{rcl} \dss\frac{\p f_{21}}{\p x^2} + \frac{\p f_{31}}{\p x^3} +  \frac{\p f_{41}}{\p x^4}   
\\ [8pt]  \dss\frac{\p f_{12}}{\p x^1} + \frac{\p f_{32}}{\p x^3} +  \frac{\p f_{42}}{\p x^4}
\\ [8pt] \dss\frac{\p f_{13}}{\p x^1} + \frac{\p f_{23}}{\p x^2} +  \frac{\p f_{43}}{\p x^4}   
\\ [8pt] \dss\frac{\p f_{14}}{\p x^1} + \frac{\p f_{24}}{\p x^2} + \frac{\p f_{34}}{\p x^3}\ea\right),\ee{lor_div}  

\n
and a comparison with  \rf{max1b} shows that \rf{max1a} is equivalent to \( \lo f = -J \) if we define \(J = (\jy, ic\rho)\) and 

\be f = (f_{ij}) =  \left( \ba{cccc} 0 & H^3 & -H^2 &  -icD^1 
\\ -H^3 & 0 & H^1 & -icD^2 
\\ H^2 & -H^1  & 0 & -icD^3 
\\ icD^1 & icD^2 & icD^3 & 0 \ea\right), \ee{fm}

\n
which is exactly the definition of $f_{ij}$ anticipated by Minkowski in \S 2-\cite{min08a}.  From it Minkowski derives the Lorentz invariants already found by Poincar\'e, but even in this case he abstains from properly referring to him.

Let us come back to what Minkowski called ``the theorem of relativity by Lorentz". As identified with a subgroup of $SO (4,\C)$, $\LCU^+$ is generated by the rotations on 2-dimensional planes around the origin. A rotation $R_\te$ by an ordinary angle $\te$ of the plane $(x^1, x^2)$ gives us easily the law of transformation of the components of $f$, once we assume the correspondence \rf{fm}, {\sl since we do know how the components of ordinary 3-dimensional vectors transform under an ordinary rotation}.\footnote{Notice that this argument is presented by Minkowski {\sl before}  the concept of vectors of kind I and II is introduced.} However a permutation of the indices like, in particular, 

\[ \left(\ba{cccc} 1 & 2 & 3 & 4 \\ 3 & 4 & 1 & 2 \ea\right) \]

 \n
 transforms \rf{max1b} into itself, and similarly it does so for \rf{max2b}. Thus by performing this permutation of the indices in the transformation law for $f$ corresponding to $R_\te$, we obtain the transformation law for $f$ corresponding also to a rotation of the $(x^3, x^4)$ plane. This rotation, however, must be done through a purely imaginary angle $i\te$, since $x^4$ is imaginary while $x^3$ is real. Now a simple computation, which Minkowski provides in \S 3, shows that such a rotation through an imaginary angle is precisely a Lorentz transformation with velocity along the $x^3$ axis. In this sense Minkowski suggests that ``the relativity theorem by Lorentz can be derived immediately, with no computations, based on the symmetry of \rf{max2a} and \rf{max2b} with respect to the indices 1,2,3,4 [...]" (\cite{min08a}, p. 59).  The basic idea, once again, is Poincar\'e's statement, cited above, that ``the Lorentz transformation is nothing but a rotation in this space around the origin, viewed as fixed'' (\cite{poi06}, p. 168). 

It must be noted, however, that it is not so obvious, before we are told (in the later \S 5-\cite{min08a}) what kind of algebraic entity $f$ is assumed to be, that its transformation law can be obtained by assuming that the `hidden' components of $\E$ and $\B$ transform under an ordinary spatial rotation just as they used to do when they were considered as {\sl bona fide} 3-dimensional vectors. Therefore Minkowski's claim that he has succeded in deriving {\sl the} transformation for the fields ``immediately with no computations'' is unwarranted.  And, after the algebraic identity of $f$ is stated, it is at the very least unclear whether we have a {\sl derivation} rather than a {\sl postulate}.    

The corresponding vector of kind II obtained by substituting in \rf{fm}

\[ ic\Dy \leadsto \B, \; \Hy \leadsto i\E /c  ,\] 

\n
as suggested  by comparing \rf{max1b} with \rf{max2b}, is

\be \left( \ba{cccc} 0 &  iE^3/c & -iE^2 /c &  -B^1 
\\ -i E^3 /c & 0 & iE^1/c & -B^2 
\\ iE^2 /c & -iE^1/c  & 0 & -B^3 
\\ B^1 & B^2 & B^3 & 0 \ea\right) = -\mu f^\ast =: F^\ast. \ee{fd_m}

\n
Therefore \rf{max2b} is equivalent to \(\lo F^\ast = 0 \). So the Maxwell equations in this formalism can be written as:

\be \left\{\ba{rcl} \lo f &=& -J  \\ [4pt] \lo F^\ast &=& 0 \ea \right., \ee{max_min}

\n
a system which appears in \S 12-\cite{min08a}, and represents an elegant and very `unified' version of the electromagnetic equations. This is the most original contribution by Minkowski, as far as the unification issue is concerned (but, needless to say, it is by no means his only contribution to relativity).

\section{Relativistic electromagnetism in modern presentation}

In modern textbooks the stress on the similarity achieved between Euclidean and Lorentzian algebra by use of the imaginary unit is normally absent (cf. e.g. \cite{sw77}, \cite{rin82}). In the following the symbols $F$ and $J$ will be given a different meaning than in the previous section. The advent of general relativity has made the attempt at formally mirroring the Euclidean geometry in Minkowski space-time neither enlightening nor computationally useful. The electromagnetic field (as expressed in the ($\E$,$\B$) format) is a 2-form $\hF$ such that in every Minkowskian coordinate system its matrix is:\footnote{We use here a superimposed dot on the equality sign to mean `is represented in a given coordinate system by'.}  

\be \hF \dot{=} (F_{ij}) =  \left( \ba{cccc} 0 & B^3 & -B^2 & E^1 
\\ -B^3 & 0 & B^1 & E^2 \\  B^2 & -B^1  & 0 & E^3
\\ -E^1 & -E^2 & -E^3  & 0 \ea\right). \ee{emt}

\n
This statement, by itself, implies the transformation formulas \rf{ft3}, with no need to derive them from the Maxwell equations. Notice that the Lorentz force law on a charge $q$ moving with 4-velocity $u$ translates into the spatial component of $q\hF (u)$. The assumption that $\hF$ is a closed form 

\be d\hF = 0 \ee{max-4D-1}

\n
is equivalent to the Gauss law for the magnetic field and the Faraday-Henry law. Let us denote by $\eta = (\eta_{ij})$ the matrix of the standard Lorentz metric in a Minkowskian coordinate system (i.e. $(\eta_{ij}) = \diag (1,1,1,-c^2)$) and let $(\eta^{ij})$ be its inverse matrix.\footnote{This means in particular that the 4-th coordinate is simply $t$.} By using the metrically equivalent 2-contravariant antisymmetric tensor:

\[ F^{ij} := \eta^{ik}\eta^{jl}F_{kl}, \]

\n
that is:

\be F \dot{=} (F^{ij}) = \left( \ba{cccc} 0 & B^3 & -B^2 & -E^1/c^2 
\\ -B^3 & 0 & B^1 & -E^2/c^2 \\  B^2 & -B^1  & 0 & -E^3/c^2
\\ E^1/c^2 & E^2/c^2 & E^3/c^2  & 0 \ea\right), \ee{emt_up}

\n
the other two Maxwell equations are recovered in the form of a single divergence equation:

\be \Di F = -\mu_0 J, \ee{max-4D-2} 

\n
where $J = (\jy, \rho)$ is the charge-current density 4-vector.  Another tensor metrically associated to $\hF$ which is needed for our purposes is the Hodge adjoint 2-form, defined intrinsically by means of the inner product on the space of 2-forms through the formula

\[ \te\we\star\hF = <\te, \hF>\vo_4, \]

\n
which is required for all 2-forms $\te$. The 4-form $\vo_4$ is the volume element, which in a Minkowskian coordinate system is

\[ \vo_4 = c dx^1 \we dx^2 \we dx^3 \we dt \]

\n
while $<\cdot, \cdot>$ is the scalar product induced on the 6-dimensional space of 2-forms (of course it is not itself Lorentzian, since, as is easy to verify, it has zero signature).  

Thus one finds that

\be \star\hF =    \left( \ba{cccc} 0 & -E^3 /c & E^2 /c & cB^1 
\\ E^3/c & 0 & -E^1 & cB^2 \\  -E^2 & E^1 & 0 & cB^3
\\ -cB^1 & -cB^2 & -cB^3  & 0 \ea\right), \ee{adj}   

\n
and a different way of writing the Maxwell equations is therefore:

\be d\hat{F} = 0, \; \de\hat{F} = \mu_0 \hat{J}, \ee{max-4D-3} 

\n
where $\hat{J}$ is the 1-form metrically equivalent to $J$, that is $\hat{J} = (\jy, -c^2\rho)$ in components, and the operator $\de$ is $\star d\star$. The potentials are `unified' in a single 1-form $A\dot{=} (\Ay, -\psi)$, where

\[ \hF = - dA, \]

\n
the two wave equations \rf{maxp} being `unified' into

\[ \Box A = -\mu_0 \hat{J}, \]

\n
subject to the Lorenz condition and to the continuity equation, which can now be expressed, respectively, as 

\[ \de A = 0, \; \de \hat{J} = 0 .\]

\n
Notice that $A$ is not uniquely defined, since one can add to it any harmonic 1-form $a$, that is any 1-form $a$ satisfying 

\[ \de a = 0, \; da = 0 .\]

From this formulation it is easy to find the two scalar invariants of the electromagnetic field for the Lorentz group, first singled out by Poincar\'e (as we have seen in \S 4.2), since we have:

\[ \hF\we\star\hF = (|\B|^2 - \frac{1}{c^2} |\E|^2)\vo, \;  \hF \we\hF = \frac{2}{c} (\E\cdot\B)\vo. \]

Overall, contemporary presentations of relativistic electromagnetism are rather close to Minkowski's approach, both in using 4-dimensional differential operators and in its axiomatic, rather than constructive, style.\footnote{A more detailed comparison of the modern and the Minkowski's formulations of electromagnetism is provided in \cite{hehl}.} However, there is something to be said in favour of giving, in agreement with Poincar\'e's approach as reconstructed above (\S 4.2.1), the retarded potential formulas a central role in order to define the fields for given charge and current densities, thus providing a substantially unambiguous proof of the transformation laws for the fields from the Maxwell equations. In the followwing subsection we shall see how the potential-based approach is helpful in clarifying a recent controversy.

\subsection{Recovering the 3-fields from the 4-fields, and a recent criticism}

If $u = e_4^\phi$ is an inertial observer of a certain Minkowski coordinate system $\phi$, then the 3-dimensional electric and magnetic fields observed by $u$ can be obtained as

\be \hF (u, \cdot) \dot{=} (\E, 0), \; \star\hF (\cdot, u) \dot{=}c (\B, 0), \ee{obs_emf}

\n
Notice that $\hF(u,\cdot)$ and $\hF(u',\cdot)$, for two different inertial observers are {\sl different 1-forms}, so by applying a Lorentz transformation to
the 4-tuple of components of \(\hF(u,\cdot)\) in $\phi$ one does {\sl not} get the 4-tuple of components of \(\hF(u',\cdot)\) in $\phi'$!  In other words, there does not exist a single 1-form corresponding to the electric field which different coordinate systems identify with different, Lorentz-related, 4-tuples of real numbers.   
 
In several articles published in the last decade (\cite{ive03}, \cite{ive05}) T. Ivezi\'c has challenged the transformation laws for the electric and magnetic fields \rf{ft3}, claiming that they ``{\sl are not} relativistically correct transformations in the 4D space-time and consequently that the usual ME [=Maxwell Equations] with $\E$ and $\B$ and the FE [= Field Equations] with $F^{ab}$ {\sl are not} physically equivalent'' (\cite{ive03}, p. 1339, italics in the original text). From the circumstance that $\hF (u,\cdot)$ and $\hF (u', \cdot)$ are in general different 4-dimensional entities, Ivesi\'c infers that the standard relativistic transformations \rf{ft3bis} of the electric and magnetic fields are incorrect. 

We think that this criticism is invalid, as it misconstrues space-time {\sl unification} as mere 4-dimensional {\sl translation} of 3-dimensional entities and equations. There is nothing wrong in the circumstance that in passing from 3 to 4 dimensions the electric and the magnetic fields turn out to lose their individuality -- after all, this is what one expects from an `unification' of 3-dimensional entities in space-time. On the other hand. the ascent from 3-vector to 4-vector applies quite naturally to the vector potential, which is welded to the scalar potential into a 4-potential.

\section{Coda. Who discovered ``Lorentz's theorem of relativity"?}

In the first decade of the 20th century the program of setting Maxwell electrodynamics on a new ground was embraced by several eminent scientists. We have examined the steps taken by the main authors during 1904-1908 towards what is now known as relativistic electrodynamics. The main conclusion of our study is that the decisive step towards a Lorentz invariant electrodynamics was the one made by Poincar\'e, although Minkowski gave the Lorentz invariance a more explicit and striking formal presentation. 

Both Poincar\'e and Einstein started with the equations to infer the transformation laws for, respectively, the potentials and the fields. We have seen that, under historically reasonable assumptions on what Poincar\'e could have taken for granted, his argument can be reconstructed as providing also the crucial {\sl uniqueness} of the transformation law -- which is tantamount to determining the algebraic nature of the $(\E, \B)$-pair. This contrasts with Einstein's argument for the necessity of the transformation laws for the fields, which is inconclusive and, more seriously, misses the direct link between the Lorentz transformation and the transformation laws for the fields (as we have seen, this is the reason Einstein has to deal with {\sl two} conformal factors, which in principle are independent). 

As to Lorentz, he  did not pretend to {\sl deduce} the transformation laws from the equations, and neither did Minkowski. 

To Minkowski's merit, it must be said that his treatment was more consistently 4-dimensional than Poincar\'e's, insofar as he introduced 4-dimensional operators to express the electrodynamical equations, and coined a new entity (the vector of kind II)  of which the ordinary electric and magnetic fields could be seen as `shadows' (to use in a different context the metaphor Minkowski famously used at the beginning of his \cite{min08b}).\footnote{Cf. Weyl: ``We are indebted to Minkowski for recognizing clearly that the fundamental equations for moving bodies are determined uniquely by the principle of relativity if Maxwell's theory for matter at rest is taken for granted. He it was, also, who formulated it in its final form ({\sl vide} \cite{min08a})'' [\cite{weyl}, p. 196].} However, not even Minkowski thought of achieving a higher degree of unification by embedding his theoretical program in the framework of the absolute calculus (i.e. tensor theory) of Ricci-Curbastro and Levi-Civita. 

Given the importance of the unification of the electric and magnetic fields in special relativity, what precedes is relevant to a serious appraisal of the respective weight of the contributions of the named authors to the creation of special relativity. It is in this connection that the evident underplaying of Poincar\'e's role by some representative mathematicians and physicists associated with the University of G\"ottingen, and not only in the first years of special relativity, appears most remarkable.\footnote{On the research and seminar activity on electromagnetism at G\"ottingen in the years just before the rise of relativity detailed information can be found in chapter 5 of \cite{pye}.}  

To mention an example from a decade later, in 1918, here is how in his famous treatise Hermann Weyl put it (\cite{weyl}, pp. 165, 173 -- the emphasis is Weyl's):

\begin{quote}
Lorentz and Einstein recognized that not only equation [\rf{maxp}] but also the {\sl whole system of electromagnetic laws for the aether has this property of invariance, namely, that these laws are the expression of invariant relations between tensors which exist in a four-dimensional affine space whose coordinates are $t, x^1, x^2, x^3$ and upon which an indefinite metrics is impressed by the form} [$-c^2 t^2 + (x^1)^2 + (x^2)^2 + (x^3)^2$].

This is the {\sc Lorentz-Einstein Theorem of Relativity}. [...]

[...] Maxwell's equations satisfy Einstein's Principle of Relativity, as was recognized even by Lorentz [...]
\end{quote}

\n
Needless to say, no such theorem was proven or even conjectured by either Lorentz or Einstein in 1904-5 -- among other reasons because, contrary to Poincar\'e,  there is no evidence that they were working with 4-dimensional space-time in mind... Indeed, Einstein does not seem initially to have found the space-time formalism particularly enlightening: in a paper written with Jakob Laub he stressed that Minkowski's approach ``makes rather great demands on the reader'', which justified the authors's offering a derivation of  his results ``in an elementary way'' (\cite{ei_lau08}, p. 532). As to Weyl, particularly remarkable in his otherwise magnificent book is the total absence of any mention of Poincar\'e's work.  

The most well-known collection of papers on the principle of relativity (\cite{ei}), first appeared in German in 1913, and translated into English in 1923 from the 4th (1922) German edition, is notorious for omitting to reproduce any excerpts from Poincar\'e's two articles on the dynamics of the electron (the whole of \cite{poi05b} and at least the first two sections of \cite{poi06} would have been an economical but fair representative sample). This omission, which from a historical point of view is simply inexcusable, can only be explained by external (i.e. extra-scientific) reasons. Such an unfair treatment is certainly not mitigated by the annotations by Sommerfeld to Minkowski's lecture (\cite{min08b}), which include two references to \cite{poi06}. The one which is relevant to us is the following (\cite{ei}, p. 96):

\begin{quote}
The invariant representation of the electromagnetic field by a ``vector of the second kind'' (or, as I proposed to call it, a ``six-vector'', a term which seems to be winning acceptance) is a particularly important part of Minkowski's view of electrodynamics. Whereas Minkowski's ideas on the vector of the first kind, or four-vector, were in part anticipated by Poincar\'e (Rend. Circ. Mat. Palermo, 21, 1906), the introduction of the six-vector is new.
\end{quote}

\n
As we have noted, while it cannot be denied that Minkowski gave the `entangled' transformation laws of electric and magnetic fields a firmer formal ground by introducing explicitly a specific algebraic entity and a 4-dimensional operator, it is nonetheless plausible that Poincar\'e disregarded this formal development simply because he did not really need it: for his purposes fixing the way components change was enough. In the same vein, one might take to task Minkowski for not mentioning that his vectors of kind I and II, with their mysteriously different transformation laws, were just particular cases of a more general algebraic entity: the Ricci-Curbastro's tensors.     
  
In the volume of his lectures devoted to electrodynamics,  Sommerfeld gives only a short reference to Poincar\'e's 1906 article, while extolling Minkowski's work, on whose shoulders he declares to stand in his treatment of the theory of relativity (\cite{som}, p. 226). He writes:
  
 \begin{quote}
 From the standpoint of the Maxwell equations the theory of relativity is obvious. A mathematician whose eyes had been trained by Klein's Erlangen program could have read from the form of the Maxwell equations its transformation group along with all its kinematical and optical consequences. [\cite{som}, p. 235]
  \end{quote}

\n
This remark is apparently oblivious of the fact that Poincar\'e's approach was itself thoroughly group-theoretic, and very much in the spirit of Klein's Erlangen program. In particular, the last section of \cite{poi06} is a {\sl tour de force} in the invariant theory of the Lorentz group, and (needless to say) the first ever on record. In his article for Einstein's 70th birthday, Sommerfeld failed to mention Poincar\'e even at a very suitable place (\cite{schi}, pp. 99-100):

\begin{quote} 
This invariance of natural laws exists in that group of motions (the uniform translations), to which Einstein [sic!], after the prior work of the great Dutchman H. A. Lorentz, has given the name of ``Lorentz-trasnformations'', although their true nature was first really grasped only by Einstein himself.
\end{quote} 

\n
Sommerfeld's belittling of Poincar\'e's contribution has surely its source in his loyalty to Minkowski's version of the story, as sketched in the introductory section of \cite{min08a}, a passage from which we have already cited a few lines:

\begin{quote}
In 1895 H. A. Lorentz [\cite{lor95}] published his theory of optical and electrical phenomena in moving bodies; this theory was based upon the atomistic representation of electricity, and on account of its great success appears to have justified the bold hypotheses, by which it has been ushered into existence. In his theory [\cite{lor95}], Lorentz proceeds from certain equations, which must hold at every point of ``Aether"; then by forming the average values over ``physically infinitely small" regions, which however contain large numbers of electrons, the equations for electro-magnetic processes in moving bodies can be successfully built up.

In particular, Lorentz's theory gives a good account of the non-existence of relative motion of the earth and the luminiferous ``Aether"; it shows that this fact is intimately connected with the covariance of the original equation, when certain simultaneous transformations of the space and time co-ordinates are effected; these transformations have therefore obtained from H. Poincar\'e [\cite{poi06}] the name of Lorentz transformations. The covariance of these fundamental equations under Lorentz transformations is a purely mathematical fact; I will call this the Theorem of Relativity; this theorem rests essentially on the form of the differential equations for the propagation of waves with the velocity of light. [...]

H. A. Lorentz has found out the ``Relativity theorem" and has created the Relativity-postulate as a hypothesis that electrons and matter suffer contractions in consequence of their motion according to a certain law.

A. Einstein [\cite{ei05}] has brought out the point very clearly, that this postulate is not an artificial hypothesis but is rather a new way of comprehending the concept of time which is forced upon us by observation of natural phenomena. [\cite{min08a}, pp. 53-5]
\end{quote}

\n
In these few lines Minkowski succeeds in the remarkable scholarly feat of: 1)  giving Lorentz the merit of having found the ``purely mathematical fact'' of the Lorentz covariance of the Maxwell equations and of having ``created the relativity-postulate''; 2) giving Einstein the priority in clarifying the link between the Lorentz transformations and a new concept of time; 3) suggesting that the main (if not the only) contribution by Poincar\'e is to have given the Lorentz transformations... their name. None of these claims is true -- and most outrageously the first one, which was explicitly rejected by Lorentz himself, as we shall see in a moment.\footnote{In his essay on ``participant histories [of relativity] in Germany'' between 1905 and 1911 Staley remarks that in \cite{min08a} Minkowski ``held Lorentz to have discovered the theorem of relativity[...]'' (\cite{sta}, p. 283), but he does not seem interested in the issue of whether this claim was fact or fiction.} What is quite clear is that Minkowski was overanxious to earn a decisive place in the historical development of the theory, and that to this end he did not care to give even such an eminent foreign colleague as Poincar\'e, whose work he followed with the utmost attention,\footnote{In 1905-1907 ``[...]  the G\"ottingen mathematical society paid attention to Poincar\'e's contributions to celestial mechanics, mathematical physics, and pure mathematics. It also appears that no other member of the mathematical society was quite as assiduous in this respect as Minkowski'' (\cite{wal06}, p. 214).} blatantly less than his due.\footnote{It is all too easy to guess, and especially instructive in our times of  reigning ``impact factors'', whether a low profile author -- like Marcolongo (\cite{marc06}) -- could have fared better with Minkowski and his disciples.} And as a matter of fact it must be added that in his famous lecture ``Space and Time'', Minkowski went to the lengths of claiming to have done ``an attack to the concept of space'' such as ``neither Lorentz nor Einstein [sic]'' (no mention of Poincar\'e, of course) had dared to do, thus absurdly pretending to have been first in realizing that the Lorentz group implies the relativity of simultaneity (\cite{min08b}, \cite{ei}, p. 83).\footnote{For a discussion with several quotations see \cite{wal99}, \S 3.4. A full examination of this issue is beyond the scope of the present paper.} A benevolent psychological interpretation of this self-inflation is contained in a letter addressed by Weyl to Minkowski's sister, Fanny, in 1947.\footnote{``Someone who contributes to a field foreign to himself is easily inclined, in the pride of also having mastered something foreign and lacking an overall view, to make an exaggerated assessment of his contribution. The lecture [i.e. \cite{min08b}] suffers also from the fact that he wanted to fix or immortalize a transitional phase in physics'' (ci. in \cite{cor97}, p. 311).}

It took no less a mathematical and academic authority than Felix Klein to insist with Wolfgang Pauli, when the latter was writing his famous article ``Relativit\"atstheorie'' for the German {\sl Encyclopedia of Mathematical Sciences}, that Poincar\'e's contribution should receive a conspicuous mention, out of respect for the historical truth (\cite{pau79}, pp. 27-8; cf. \cite{enz}, pp. 28-9). This recommendation, as far as the unification issue is concerned, resulted in the following passage being inserted (\cite{pau}, p. 78):  

\begin{quote}
In his paper of 1904, Lorentz [\cite{lor04}] came very near to proving the covariance of [Maxwell equations] under the relativistic transformation group. The complete proof was given, independently, by Poincar\'e [\cite{poi06}] and Einstein [\cite{ei05}]. The four-dimensional formulation is due to Minkowski [\cite{min15, min08a, min08b}], who first stressed the concept of a ``surface'' tensor, as we would call it now. 
\end{quote}

\n
This is not completely wide of the mark, but neither is completely satisfactory, although in a footnote Pauli had already specified that ``As a precursor of Minkowski one should mention Poincar\'e [...]'' for having introduced the imaginary time coordinate and for having ``combined, and interpreted as point coordinates in $\R^4$, those quantities which we now call vector components. Furthermore, the invariant interval plays a role [in fact quite an important one!] in his considerations'' (\cite{pau}, p. 21, fn. 54). 

Finally let us come to Lorentz's version of the historical facts. In his 1915 tribute to Poincar\'e Lorentz made the following remarks to explain why he had got the formulas for the charge and current densities wrong (\cite{lor15}, p. 297-8; italics in the original):

\begin{quote}
The [right] formulas [for the charge and current densities] are not to be found in my paper of 1904. The reason is that I had not thought of the straight path that leads to them, and this depends on the fact that I had the idea that there is an essential difference between the systems $[\phi]$ and $[\phi']$. In one of them one uses -- such was my thought -- coordinate axes with a fixed position in the aether and what one can call the `true' time; in the other system, on the contrary, one would have to do with simple auxiliary quantities, which are introduced by just a mathematical artifice. In particular, variable $t'$ could not be called `time' in the same sense as variable $t$. 

In this frame of mind I have not thought of describing the phenomena in system $[\phi]$ {\sl in exactly the same manner} as in system $[\phi]$ [...] Later I could see in Poincar\'e's paper that by proceeding more systematically I could have achieved a still greater simplification. Not noticing it, I failed to obtain the exact invariance of the equations; my formulas remained encumbered with certain terms which should have disappeared. These terms were too small to have a sensible influence in the phenomena and therefore I could explain the independence of the Earth's motion which the observations had shown, but I have not established the principle of relativity as rigorously and exactly true. 

Poincar\'e, on the contrary, has obtained a perfect invariance of the equations of electrodynamics, and he has formulated the `postulate of relativity', terms which he has been the first to use. In fact, placing himself from the standpoint which I had missed, he has found the [right] formulas. Let us add that in correcting the imperfections of my work, he has never rebuked me for it.  
\end{quote} 

\n
In view of our analysis, this authoritative statement is, in particular, an accurate assessment of Poincar\'e's contribution to the unification of the electric and magnetic fields.

\end{document}